\ifdefined\pdfoutput
  \pdfoutput=1
\fi
\ifdefined\pdfminorversion
\fi
\documentclass[11pt]{article}

\usepackage[margin=1in]{geometry}
\usepackage{authblk}
\usepackage{graphicx}
\usepackage{subfigure}
\usepackage{framed,multirow}
\usepackage{amssymb}
\usepackage{amsmath}
\usepackage{latexsym}
\usepackage[numbers,sort&compress]{natbib}
\usepackage{url}
\usepackage{xcolor}
\usepackage[hidelinks]{hyperref}

\definecolor{newcolor}{rgb}{.8,.349,.1}

\title{A second-order diffusive-interface immersed boundary method for incompressible flow with phase change and moving interfaces}

\author[1,2,3]{Wenyuan Chen\thanks{Corresponding authors: \texttt{97wychen@pku.edu.cn}, \texttt{yantao.yang@pku.edu.cn}}}
\author[1]{Yantao Yang}

\affil[1]{State Key Laboratory for Turbulence and Complex Systems, Department of Mechanics and Engineering Science, College of Engineering, Peking University, Beijing 100871, China}
\affil[2]{Department of Mechatronics Engineering, Morgan State University, Baltimore, MD 21251, USA}
\affil[3]{Department of Mechanical Engineering, Johns Hopkins University, Baltimore, MD 21218, USA}

\date{}

\begin{document}

\maketitle

\begin{abstract}
Accurately resolving interfacial gradients is critical for simulating two-phase flows, particularly those involving phase transitions or active matter. The traditional diffuse-interface immersed boundary methods (IBMs) are highly efficient for such problems, but they typically suffer from a reduction to first-order accuracy near the phase-changing boundaries. We clarify that the main reason is the local derivative discontinuities. Here, we propose a smooth extension strategy to restore formal second-order spatial accuracy. By extrapolating the scalar field across the interface, the method structurally ensures derivative continuity. To preserve the divergence-free condition in incompressible fluid solvers, this smooth extension is applied exclusively to the scalar transport equations. The velocity field retains the standard diffuse-interface treatment. The proposed framework is systematically validated against classical phase-change benchmarks, specifically one-dimensional evaporation and boiling problems. Additionally, the method is applied to the spontaneous autophoretic motion of isotropic particles. The numerical results confirm the capability of our method in resolving the complex multi-physics boundary couplings.
\end{abstract}



\section{Introduction}
Multiphase flows, particularly those involving phase transitions, involve complex momentum, energy, and mass transfer across various spatiotemporal scales. Accurate simulation of these phenomena is an ongoing challenge in computational fluid dynamics, where the immersed boundary method (IBM) serves as a practical tool for handling these multiphase systems. Over the past decades, IBM, originally pioneered by Peskin~\cite{peskin1972}, has become a well-established framework for resolving moving boundaries on uniform Cartesian grids~\cite{mittal2005,wang2017immersed}. Recently, the application of IBM in multiphase transport has expanded steadily~\cite{xiao2022immersed}. For example, developments have integrated IBM with Level Set and Volume of Fluid (VOF) frameworks to track topological evolutions~\cite{souza2022multi} and moving contact lines in three-dimensional multiphase systems~\cite{yan2024improved}.

In phase-change multiphase flows, the deformation and evolution of the phase interface depend heavily on localized physical phenomena, such as boundary heat, mass, and reaction fluxes. Consequently, numerical methods generally require higher resolution near the boundary to capture these gradients accurately. Furthermore, solid-liquid phase change materials and high-density-ratio multiphase flows often present steep gradient discontinuities at the interface, which can lead to numerical diffusion~\cite{jin2022combined,liu2022enthalpy}. Generally, sharp-interface immersed boundary methods are frequently applied to these problems, as they often achieve higher local accuracy than diffuse-interface approaches~\cite{badalassi2003computation,fadlun2000,seo2011sharp}. While sharp-interface methods offer distinct accuracy advantages, their geometric tracking requirements and computational overhead can introduce additional considerations for complex engineering applications~\cite{uhlmann2005,kempe2012}. For example, ensuring smooth transitions with the ghost-cell method in moving boundary problems requires careful treatment to minimize numerical oscillations~\cite{jost2021direct}. Similarly, the cut-cell method relies on detailed information about how the boundary intersects local cells, which can increase the algorithmic complexity for highly deformable geometries~\cite{malan2021geometric}. Additionally, the immersed interface method involves formulating explicit jump conditions across the interface~\cite{li1998immersed,lai2000,griffith2005order}, which introduces further analytical complexity in multiphase applications.

Conversely, while the diffuse-interface method is straightforward to implement, numerically stable, and computationally efficient, it typically yields lower local accuracy. Because the singular delta function is regularized, traditional diffuse-interface methods are generally restricted to first-order accuracy near the immersed boundary. This local degradation introduces numerical diffusion across phase boundaries, which compromises the overall simulation precision~\cite{treeratanaphitak2023diffuse}. To resolve this issue, we propose a smooth extension strategy for the immersed boundary method that restores second-order boundary accuracy by eliminating derivative discontinuities at the interface. This approach achieves formal second-order spatial resolution while preserving the computational efficiency of the diffuse-interface framework. To validate the method, we test it against classical two-phase phase-change benchmarks, including the one-dimensional evaporation and boiling problems, alongside the spontaneous autophoretic motion of active particles~\cite{malan2021geometric, shao2018computational, michelin2013spontaneous}.

Recent one-sided diffuse-interface methods have achieved higher-order accuracy in compressible flows~\cite{mao2026explicit}. However, these approaches require explicit inside-outside phase discrimination and often rely on computationally expensive iterative matrix solvers. Another strategy which is recently reported in Zhu et al.~\cite{zhu2024boundary} is a two-inner-probe method to enforce Neumann boundary conditions. The approach identifies one fluid-side probe and uses linear extrapolation to prescribe the scalar values at two solid-side probes. They noted that this linear extrapolation improves the continuity of the concentration field near the boundary. In this study, we propose a different method to improve the accuracy of scalar field at immersed boundary with phase-changing process. We first clarify the key source of local accuracy degradation through a rigorous truncation error analysis, then construct a smooth extension method for the scalar field. Similar smooth extension has been discussed before~\cite{beyer1992analysis, stein2016immersed}. The method is also incorporated with the non-iterative and low error scheme developed in our previous work~\cite{chen2023}. 

It should be pointed out that, for incompressible flows, directly modifying the velocity field near the boundary interferes with the projection step and violates the divergence-free condition, causing numerical instability. Our framework therefore adopts a hybrid treatment. The smooth extension is applied exclusively to the scalar fields (such as temperature or concentration) that dictate the interfacial dynamics and phase-change rates. Concurrently, the velocity field retains the standard diffuse-interface formulation. We validate the method using classical phase-change benchmarks and simulations of spontaneous autophoretic motion, demonstrating improved boundary accuracy in resolving complex multi-physics couplings.

The remainder of this paper is organized as follows. Section 2 presents the governing equations and the baseline numerical method. Section 3 analyzes the mathematical sources of boundary errors and introduces the proposed smooth extension strategy to restore second-order accuracy. Section 4 presents numerical validations against classical phase-change and autophoretic motion benchmarks. Finally, conclusions are drawn in Section 5.

\section{Governing equations and the baseline numerical method}

This section presents the governing equations and the numerical scheme of the flow solver. We outline now the baseline immersed boundary (IB) method, which serves as the foundation for the improved framework developed in this study.

\subsection{Governing equations and flow solver}

We confine ourselves to the incompressible flows coupled with a scalar transport component. Denoting the velocity vector by $\mathbf{u}=(u,v,w)$ and the scalar field by $S$, the incompressible Navier-Stokes equations and the scalar advection-diffusion equation are given as:
\begin{subequations}\label{eq:govern}
	\begin{eqnarray}
		\nabla \cdot \mathbf{u} &=& 0, \label{eq:continu} \\
		\partial_t \mathbf{u} + \mathbf{u}\cdot\nabla\mathbf{u} &=& -\frac{1}{\rho_L}\nabla p + \nu \nabla^2 \mathbf{u} + \mathbf{f}_u, \label{eq:ns} \\
		\partial_t S + \mathbf{u}\cdot\nabla S &=& \kappa \nabla^2 S + f_S. \label{eq:scalar}
	\end{eqnarray}
\end{subequations}
Here, $\rho_L$ is the fluid density, $\nu$ the kinematic viscosity, and $\kappa$ the molecular diffusivity, respectively. The virtual forcing terms $\mathbf{f}_u$ and $f_S$ are introduced to enforce the desired boundary conditions on the velocity and scalar fields, respectively. Accurately determining these two forcing terms is the core task of the IB method and will be discussed in detail in subsequent sections. The Reynolds number is defined as $Re=UL/\nu$, using the characteristic velocity $U$ and characteristic length $L$. 

The governing equations \eqref{eq:govern} are solved using a fractional-step method, which has been extensively validated for wall-bounded and convective turbulence~\cite{verzicco1996,ostilla2015multiple}. Spatial discretization is performed using a conservative second-order central difference scheme on a staggered grid. Temporal integration is carried out via a Runge-Kutta (RK) scheme, where the non-linear terms are treated explicitly using an Adams-Bashforth approach, and the viscous and diffusive terms are handled implicitly via the Crank-Nicolson scheme. Because a Cartesian domain is employed, the large sparse linear system resulting from the implicit treatment is efficiently solved using an approximate factorization method. The pressure Poisson equation is solved utilizing a fast-Fourier-transform (FFT) algorithm. Comprehensive details regarding the numerical scheme are available in~\cite{verzicco1996, ostilla2015multiple}.

\subsection{The baseline IB method}

The baseline IB framework used in this study is based on the Moving-Least-Squares (MLS) IB method proposed in~\cite{vanella2009}. In the MLS-IB method, the immersed boundary is represented by a set of Lagrangian markers with coordinates denoted by $\mathbf{X}$, while the Eulerian grid coordinates are denoted by $\mathbf{x}$. During numerical integration, flow variables at the Lagrangian markers are interpolated from the Eulerian field, and the computed IB forcing is subsequently spread back from the markers to the surrounding Eulerian grid. The distinguishing feature of the MLS-IB method is the use of a unified MLS transfer function $\Phi(\mathbf{x},\mathbf{X})$ for both interpolation and spreading. The exact formulation of $\Phi$ is detailed in~\cite{vanella2009, vanella2020}.

In a previous study, we improved the original MLS-IB method by introducing a volume-correction coefficient applied to the Eulerian IB forcing after the standard spreading operation~\cite{chen2023}. The rationale is that because multiple Eulerian grid points simultaneously contribute to enforcing the boundary condition at several Lagrangian markers, the actual effective force exerted on the Eulerian grid often deviates from the desired analytical value, introducing local errors. While iterative procedures or implicit formulations can mitigate this~\cite{kempe2012,wang2011}, they inevitably increase computational overhead. Our previous work demonstrated that the interpolation error follows a narrow, single-peak distribution; thus, applying a single explicit correction factor $Z$ reduces the error to a level comparable to iterative methods~\cite{chen2023}. 

Given the velocity field $\mathbf{u}^n$ and pressure field $p^n$ at the $n$-th time step, the updated fields $\mathbf{u}^{n+1}$ and $p^{n+1}$ are computed using a three-substep RK method. For each RK substep (indexed by $i$), the fractional-step procedure is executed as follows:
\begin{subequations}
	\begin{eqnarray}
		\tilde{\mathbf{u}}: & \quad & \frac{\tilde{\mathbf{u}}-\mathbf{u}^n}{\Delta t}
		= -\alpha_i\nabla p^n + \gamma_i \mathbf{H}^n + \rho_i \mathbf{H}^{n-1} 
		+ \frac{\alpha_i}{2 Re}\nabla^2(\tilde{\mathbf{u}}+\mathbf{u}^n), \\
		\mathbf{U}^{l}: & \quad & \mathbf{U}^{l}\left(\mathbf{X}^{l}\right) 
		= \sum_{k=1}^{ne} \tilde{\mathbf{u}}_k \,\Phi_{k}^{l}(\mathbf{x}_k,\mathbf{X}^l),  \\
		\mathbf{F}\left(\mathbf{X}^{l}\right): & \quad & \mathbf{F}\left(\mathbf{X}^{l}\right) 
		= \frac{\mathbf{U}^{d}-\mathbf{U}^{l}}{\Delta t},  \\
		\mathbf{f}\left(\mathbf{x}_k\right): & \quad & \mathbf{f}\left(\mathbf{x}_k\right)
		= Z \sum_{l=1}^{nl} c_l \, \Phi_{k}^{l}(\mathbf{x}_k,\mathbf{X}^l) \, \mathbf{F} \left(\mathbf{X}^{l}\right), \\
		\mathbf{u}^*: & \quad & \mathbf{u}^* = \tilde{\mathbf{u}} + \Delta t\, \mathbf{f}, \\
		\psi: & \quad & \nabla^2 \psi = \frac{\nabla \cdot \mathbf{u}^*}{\alpha_i \Delta t}, \\
		\mathbf{u}^{n+1}: & \quad & \mathbf{u}^{n+1} = \mathbf{u}^* - \alpha_i\Delta t \nabla\psi, \\
		p^{n+1}: & \quad & p^{n+1} = p^n + \psi - \frac{\alpha_i\Delta t}{2 Re}\nabla^2 \psi.
	\end{eqnarray}	
\end{subequations}
Here, $\alpha_i$, $\gamma_i$, and $\rho_i$ are the standard integration coefficients for the RK scheme~\cite{verzicco1996}. $\mathbf{U}^d$ is the desired velocity at the Lagrangian markers dictated by the physical boundary conditions. The subscript $k$ denotes values at the Eulerian grid $\mathbf{x}_k$, while the superscript $l$ denotes values at the Lagrangian marker $\mathbf{X}^l$. The variables $ne$ and $nl$ represent the number of Eulerian grids involved in interpolating the marker $\mathbf{X}^l$ and the number of Lagrangian markers contributing to the forcing at grid $\mathbf{x}_k$, respectively. $c_l$ is the MLS normalization factor~\cite{vanella2009,chen2023}. The universal correction factor $Z$ is calculated explicitly as:
\begin{equation}\label{eq:zcoef}
	Z = 
	\frac{ 
		\sum_{l=1}^{NL}\left[\sum_{k=1}^{ne}\sum_{m=1}^{nk} 
		c_m\Phi_{k}^{l}\Phi_{k}^{m} \mathbf{F}(\mathbf{X}^{m})\cdot \mathbf{F}(\mathbf{X}^l)\right]
	}
	{ 
		\sum_{l=1}^{NL}\left[\sum_{k=1}^{ne}\sum_{m=1}^{nk}
		c_m\Phi_{k}^{l}\Phi_{k}^{m} \mathbf{F}(\mathbf{X}^{m})\right]^2
	},
\end{equation}
where $NL$ is the total number of Lagrangian markers. For the scalar field, an identical interpolation and spreading procedure is applied.

\section{Improved treatment for the scalar field at the immersed boundary}

This section discusses the fundamental mathematical sources of numerical error at the immersed boundary and presents the smooth extension methodology specifically designed to elevate the scalar boundary condition accuracy to second order.

\subsection{General discussion on boundary errors}

In diffuse-interface IB methods, the virtual body force added to the governing equations fundamentally acts as a regularized Dirac delta function centered at the immersed boundary. This singularity restricts the flow quantities to $C_0$ continuity, meaning spatial derivatives across the boundary are inherently discontinuous. This is illustrated by the simplified one-dimensional second-order model:
\begin{equation}
	u_{xx} = \delta(x-x_0),
\end{equation}
where $\delta(x)$ is the Dirac delta function. Integrating this equation across an infinitesimal neighborhood around $x_0$ yields the gradient jump condition:
\begin{equation}
	u_{x}(x_0^+) - u_{x}(x_0^-) = 1.
\end{equation}

This discontinuity in the first-order spatial derivative introduces systematic truncation errors during interpolation. Following an analysis similar to Beyer and LeVeque~\cite{beyer1992analysis}, we expand the general interpolation scheme using a Taylor series around $x_0$. For a function with discontinuous derivatives, the value can be expressed as:
\begin{equation}\label{eq:discontinuousfx}
	f(x) = f(x_0) + \sum_{n=1}^{\infty}\dfrac{f^{(n)}(x_0^+)}{n!}(x-x_0)^n 
	+ \sum_{n=1}^{\infty}H(x-x_0)\dfrac{[f^{(n)}]}{n!}(x-x_0)^n,
\end{equation}
where $[f^{(n)}] = f^{(n)}(x_0^+) - f^{(n)}(x_0^-)$ represents the jump in the $n$-th derivative, and $H(x-x_0)$ is the Heaviside step function. 

If interpolation is performed over a local stencil $\{x_m\}$ to evaluate the value at $x_0$:
\begin{equation}\label{eq:fstar}
	f^*(x_0) = \sum_{m}f(x_m)\phi(x_m),
\end{equation}
the resultant interpolation error $I_{error} = f(x_0)-f^*(x_0)$ expands to:
\begin{eqnarray}\label{eq:completeIerror}
	I_{error} &=& f(x_0) - f(x_0)\sum_{m}\phi(x_m) \nonumber\\
	&-&\sum_{n=1}^{\infty}\dfrac{f^{(n)}(x_0^+)}{n!}\sum_{m}\phi(x_m)(x_m-x_0)^n \nonumber\\
	&-&\sum_{n=1}^{\infty}\dfrac{[f^{(n)}]}{n!}\sum_{m}H(x_m-x_0)\phi(x_m)(x_m-x_0)^n.
\end{eqnarray}
Consequently, to achieve $k$-th order spatial accuracy, the interpolation weights $\phi(x_m)$ must satisfy:
\begin{eqnarray}
	&& \sum_{m}\phi(x_m)  = 1, \label{eq:unitcondition} \\
	&& \sum_{m}\phi(x_m)(x_m-x_0)^n = 0, \quad n=1,2,...,k, \label{eq:acurrancycondition} \\
	&& \sum_{m}H(x_m-x_0)\phi(x_m)(x_m-x_0)^n = 0, \quad n=1,2,...,k. \label{eq:additioncondition}
\end{eqnarray}
While the normalization condition \eqref{eq:unitcondition} and the standard moment condition \eqref{eq:acurrancycondition} are routinely satisfied, the unilateral moment condition \eqref{eq:additioncondition} is required exclusively when $[f^{(n)}]\neq0$. Constructing interpolation kernels that satisfy Eq.~\eqref{eq:additioncondition} typically results in non-positive definite functions~\cite{beyer1992analysis}, which provoke severe numerical oscillations and instabilities near the boundary.

\subsection{The smooth extension strategy for the scalar field}

Instead of designing complicated interpolation stencils that degrade numerical stability, we can directly decrease the derivative jump ($[f^{(n)}]=0$). For fluid-solid systems where the flow inside the solid is irrelevant, this is accomplished by smoothly extending the fluid field into the solid domain.

Several studies have utilized this extension concept. For example, Stein et al.~\cite{stein2016immersed} smoothly extended PDE solutions across arbitrary boundaries using a Fourier spectral method. More recently, Zhu et al.~\cite{zhu2024boundary} enforced Neumann conditions by applying linear extrapolation at two auxiliary layers inside the solid. In the present work, we achieve the smooth extension through explicit immersed boundary forcing.

It should be noted that the underlying philosophy of this approach shares conceptual similarities with sharp-interface immersed boundary methods~\cite{ fadlun2000,mittal2008versatile} and cut-cell techniques~\cite{udaykumar2001sharp}. In those frameworks, local second-order accuracy is achieved by explicitly reconstructing the solution near the interface using specialized asymmetric interpolation stencils or geometric grid-cutting operations. However, such methods require rigorously tracking the exact interface topology, strictly distinguishing the fluid and solid nodes, and performing complex geometric intersections. These geometric operations become computationally expensive and difficult to implement robustly for moving boundaries or problems involving topological changes, such as phase-change phenomena.

In contrast, our approach entirely circumvents these geometric complexities while retaining the algorithmic simplicity of the diffuse-interface framework. Instead of clipping the background grid or searching for asymmetric stencils, we structurally extend a smooth forcing zone across the boundary into the solid domain. By applying these extrapolated virtual forces, the standard, symmetric spreading kernels can naturally resolve the sharp boundary condition with second-order accuracy. Consequently, the method maintains the robustness and efficiency of diffuse-interface IBMs without compromising the interfacial precision.

Although the above error analysis applies to any variable, extending the velocity field poses a specific problem for incompressible flows. Modifying the velocity field inside the solid domain introduces artificial mass sources. In a fractional-step solver, this local divergence directly interferes with the pressure Poisson equation and causes numerical instability. Consequently, we restrict the smooth extension treatment to the scalar field.

\begin{figure}[htbp]
	\centering
	\includegraphics[width=0.45\textwidth]{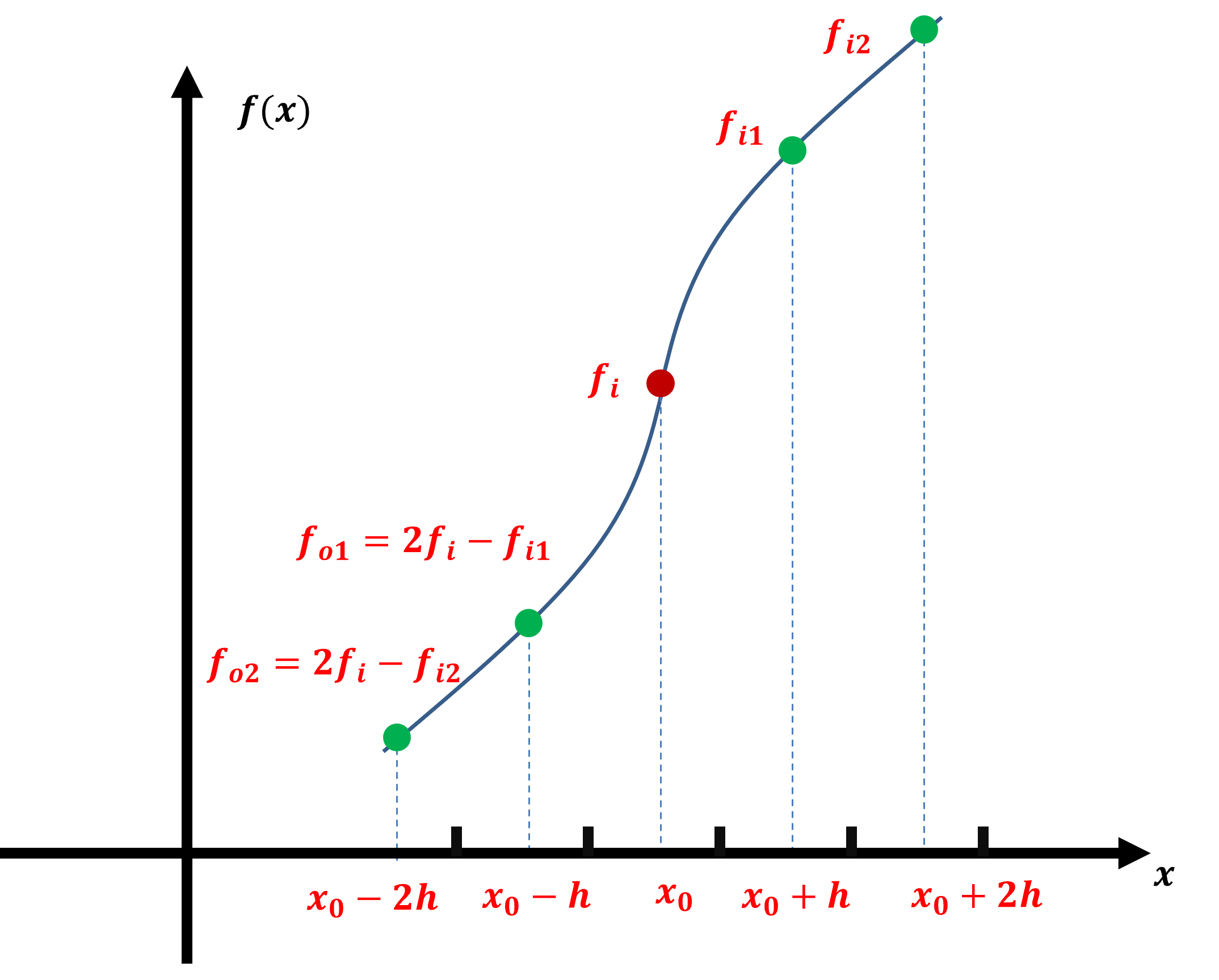}
	\includegraphics[width=0.45\textwidth]{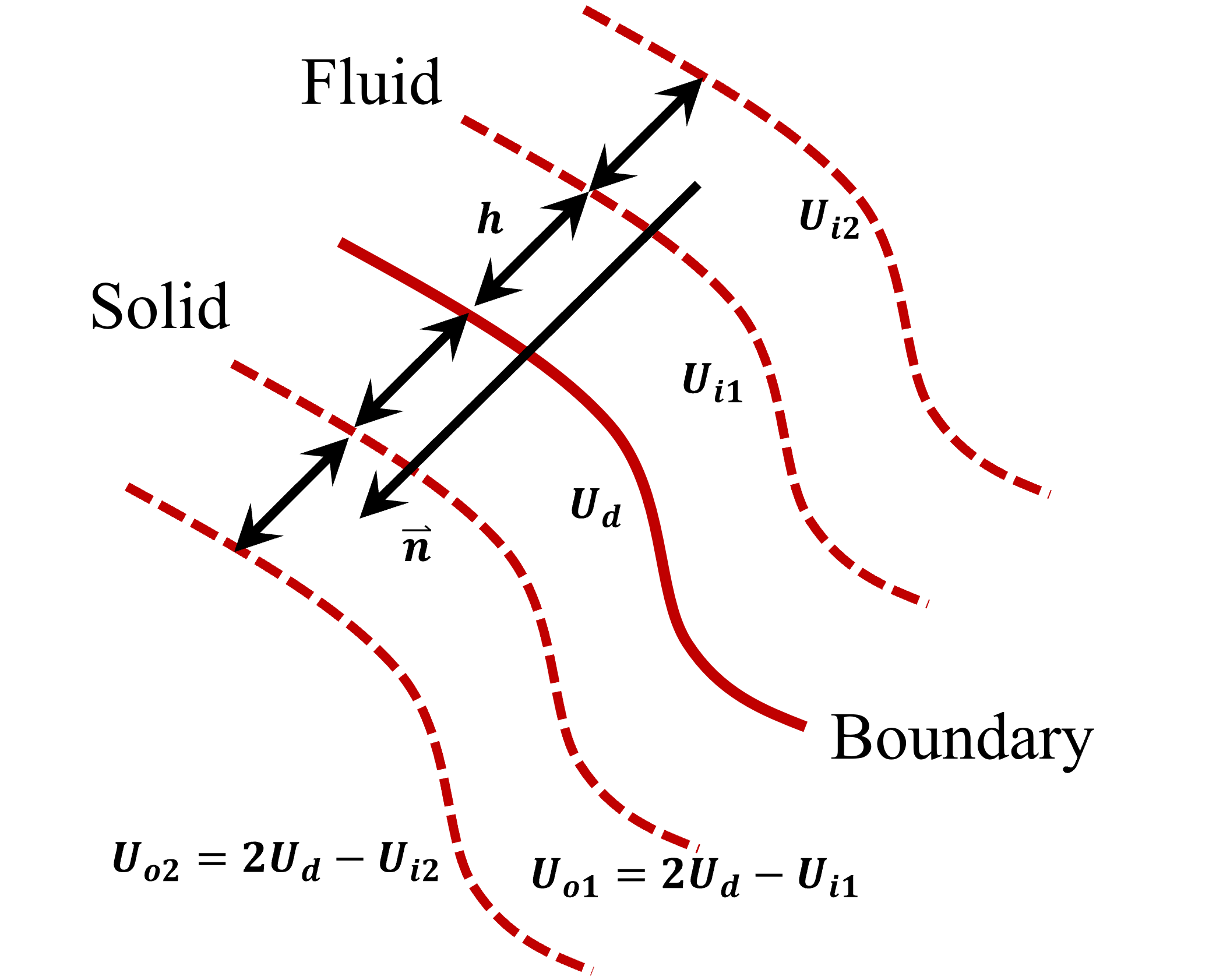}
	\caption{(a) One-dimensional schematic illustrating the reconstruction of external boundary values (auxiliary layers) to enforce derivative continuity at the immersed boundary. (b) Schematic diagram of the smooth extension implementation in a two-dimensional computational domain showing the normal extrapolation.}
	\label{fig:smooth_extension}
\end{figure}

To achieve second-order accuracy ($[S^{(1)}]=0$), the first-order spatial derivative must be continuous across the phase interface. We enforce this condition by introducing auxiliary Lagrangian markers on the solid side, oriented along the outward normal $\mathbf{n}$. The geometric configurations for locating these markers and performing the necessary interpolations in one- and two-dimensional cases are illustrated in Fig.~\ref{fig:smooth_extension}(a) and (b), respectively.

For a Dirichlet condition with a prescribed boundary value $S_b$, three layers of Lagrangian markers are actively controlled: the physical boundary itself, along with two auxiliary surfaces, denoted as $O1$ and $O2$, placed at normal distances of $h$ and $2h$ into the solid domain. Their target values are explicitly determined by mirroring the corresponding fluid-side points ($I1$ and $I2$), yielding $S_{O1}=2S_b-S_{I1}$ and $S_{O2}=2S_b-S_{I2}$. Virtual body forces are then applied to enforce the scalar field values at the physical boundary and these two auxiliary layers.

When a Neumann condition with a prescribed normal gradient $(\partial_n S)_b$ is applied, the exact scalar value at the boundary is not known a priori. To maintain gradient continuity, we only need to control two auxiliary surfaces ($O1$ and $O2$) located at normal distances of $h$ and $2h$ inside the solid domain. Using their respective mirror points ($I1$ and $I2$) in the fluid region, the target scalar values at these auxiliary surfaces are extrapolated as $S_{O1} = S_{I1} + 2h(\partial_n S)_b$ and $S_{O2} = S_{I2} + 4h(\partial_n S)_b$. Enforcing these extrapolated values structurally guarantees gradient continuity across the phase interface without actively forcing the physical boundary itself.

\subsection{Detailed procedure of the proposed IB method}

Before evaluating the immersed boundary forces, the geometric locations of the auxiliary layers and their corresponding fluid-side mirror points must be rigorously constructed. The physical phase boundary is discretized by a set of Lagrangian markers located at the vertices of triangular elements. Along this normal direction, the discrete points for the auxiliary layers ($\mathbf{X}^{O1}$, $\mathbf{X}^{O2}$) and the fluid-side mirrors ($\mathbf{X}^{I1}$, $\mathbf{X}^{I2}$) are projected at distances of $h$ and $2h$, respectively. To prevent ray-crossing artifacts in regions of extreme surface curvature, the projection distance $h$ is typically chosen to be comparable to the local Eulerian grid spacing $\Delta x$, assuming the interfacial geometry is adequately resolved by the background mesh.

With the geometry defined, the numerical procedure within a single RK sub-step proceeds as follows. Following Kim et al.~\cite{kim2001}, an explicit Euler prediction is first evaluated for the scalar field to determine the necessary unforced intermediate state. The explicit time-advancement for the predicted scalar field $S_E$ is given by:
\begin{equation}
	\frac{S_E-S^n}{\Delta t} = \gamma_i H_s^n + \rho_i H_s^{n-1} + \alpha_i\kappa\nabla^2 S^n.
\end{equation}

Our numerical tests demonstrate that the sequence of applying the boundary forces critically impacts the overall accuracy. To systematically enforce the smooth extension, the operations must proceed sequentially from the outermost auxiliary layer inward toward the fluid domain. Taking the Dirichlet condition as an example, the calculation initiates at the outermost layer ($O2$). The intermediate scalar values at both the auxiliary markers and their fluid-side mirrors are interpolated from the predicted Eulerian field:
\begin{eqnarray}
	S^{I2}\left(\mathbf{X}^{I2}\right) &=& \sum_{k=1}^{ne} S_{E,k} \,\Phi_{k}^{I2}(\mathbf{x}_{k},\mathbf{X}^{I2}), \\
	S^{O2}\left(\mathbf{X}^{O2}\right) &=& \sum_{k=1}^{ne} S_{E,k} \,\Phi_{k}^{O2}(\mathbf{x}_{k},\mathbf{X}^{O2}).
\end{eqnarray}
Based on the smooth extension relation, the target scalar value at this outer layer is explicitly determined as $S_d^{O2} = 2S_b - S^{I2}$. The required Lagrangian force $F^{O2}$ and the corresponding spread Eulerian force $f_{O2}^{n+1/2}$ are then computed, incorporating the volume-correction coefficient $Z$:
\begin{eqnarray}
	F^{O2} &=& \frac{S_d^{O2}-S^{O2}}{\Delta t},  \\
	f_{O2}^{n+1/2}(\mathbf{x}_k) &=& Z \sum_{l=1}^{nl} c_{l} \, \Phi_{k}^{l}(\mathbf{x}_{k},\mathbf{X}^{O2,l}) \, F^{O2} \left(\mathbf{X}^{O2,l}\right).
\end{eqnarray}
Crucially, the predicted Eulerian scalar field is immediately updated with this local forcing before proceeding to the next layer:
\begin{equation}
	S_E \leftarrow S_E + \Delta t\, f_{O2}^{n+1/2}.
\end{equation}

Using this updated $S_E$ field, the identical interpolation, forcing, and spreading procedure is repeated for the inner auxiliary boundary $O1$ to obtain $f_{O1}^{n+1/2}$. After a second update to the intermediate scalar field, the process is executed one final time at the actual physical boundary $b$ to evaluate $f_{b}^{n+1/2}$.

The total IB force applied to the scalar transport equation is the exact superposition of these sequentially derived components: $f^{n+1/2} = f_{O2}^{n+1/2} + f_{O1}^{n+1/2} + f_{b}^{n+1/2}$. This sequential updating strategy ensures that the enforcement of boundary conditions at the inner layers properly accounts for the mathematical influence of the extrapolated outer layers, thereby maintaining the desired gradient continuity.

\subsection{Location update of the phase-changing immersed boundary}

For cases involving phase transition, the local topology of the immersed boundary deforms dynamically. The boundary is discretized into triangular elements, requiring the vertices to be updated at every time step. Given the normal moving velocity $\mathbf{V}_n$ of the center of an element $n$, the displacement of the element center is:
\begin{equation}
	\mathbf{d}_n = \alpha_i\mathbf{V}_n\Delta t.
\end{equation}
Because multiple triangular elements share a single vertex, the displacement of each vertex must be volume-conserving. This is achieved using an area-weighted average of the displacements of all connected elements:
\begin{equation}
	\mathbf{D}_{vertex} = \frac{\sum_{n=1}^{NE} \mathbf{d}_n A_n}{\sum_{i=1}^{NE} A_i},
\end{equation}
where $NE$ is the number of elements sharing the specific vertex, and $A_i$ is the area of the $i$-th connected element.

\section{Validations and applications of the improved method}

\subsection{The one-dimensional Stefan problem}

\begin{figure}[htbp]
	\centering
	\includegraphics[width=0.4\textwidth]{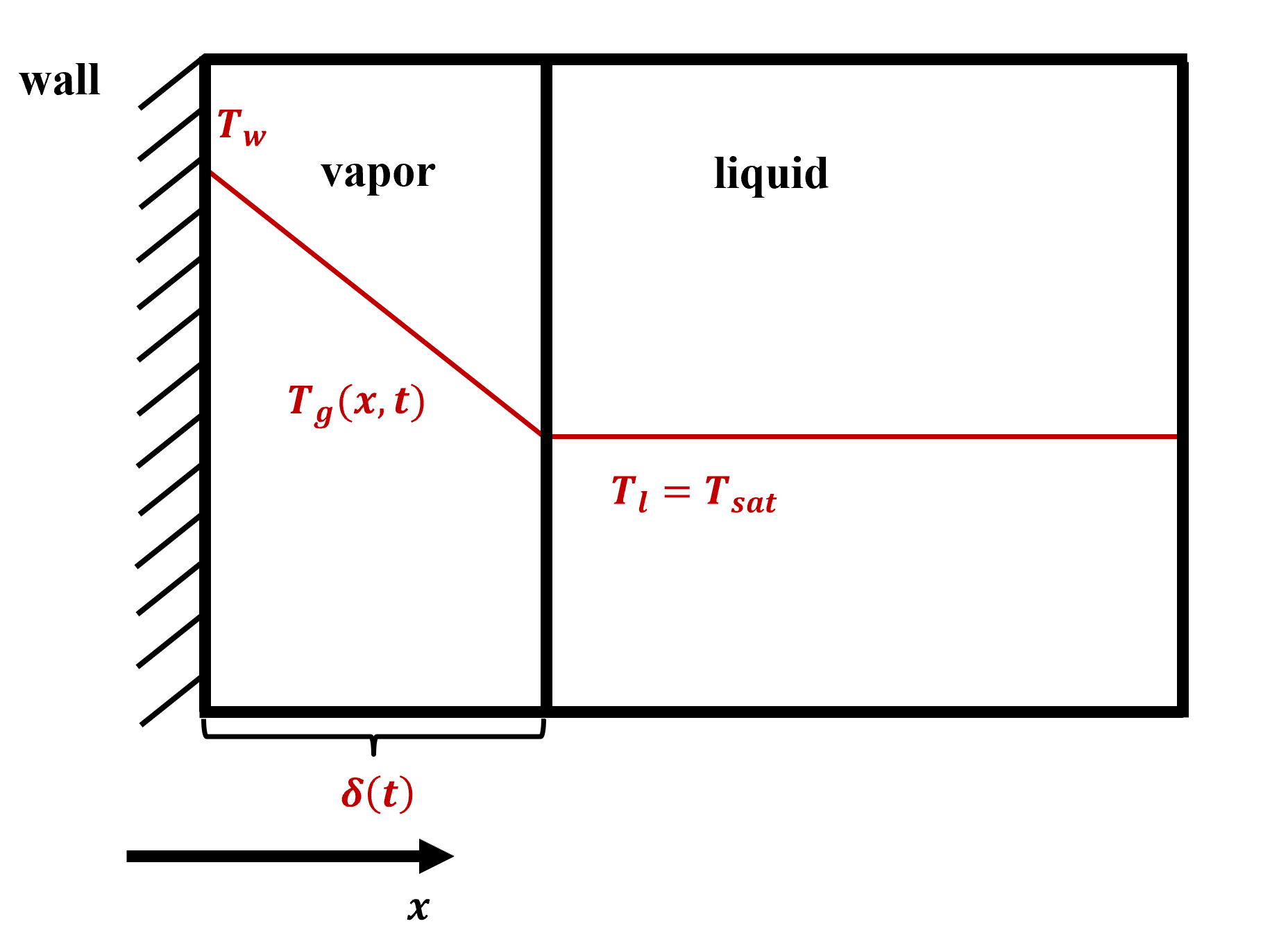}%
	\caption{Configuration of the one-dimensional evaporation (Stefan) problem.}
	\label{fig:stefan}
\end{figure}

The one-dimensional Stefan problem is a standard benchmark for validating numerical methods involving phase-change interfaces~\cite{welch2000volume,hardt2008,sato2013sharp}. As illustrated in Fig.~\ref{fig:stefan}, the domain initially consists of a liquid at rest occupying the half-space $x > 0$, bounded by a solid wall at $x=0$. The liquid is maintained at the saturation temperature $T_{sat}$, while the wall is kept at a constant superheated temperature $T_w > T_{sat}$. Evaporation occurs at the wall, driving the growth of a stationary vapor layer that displaces the liquid to the right. Let $\delta(t)$ denote the position of the vapor-liquid interface. The transient heat conduction within the vapor layer is subjected to Dirichlet boundary conditions at both the heated wall and the phase interface, governed by:
\begin{eqnarray}
	\frac{\partial T}{\partial t} &=& \alpha_g \frac{\partial^2 T}{\partial x^2}, \qquad 0 < x < \delta(t), \\
	T(0, t) &=& T_w, \\
	T(\delta(t), t) &=& T_{sat}.
\end{eqnarray}
The phase-change dynamics at the interface are coupled to the local temperature gradient through the Stefan condition, which dictates the interface moving velocity $v_\gamma$:
\begin{eqnarray}
	\rho_g h_{lg} v_\gamma &=& -k_g \left. \frac{\partial T}{\partial x} \right|_{x=\delta(t)}, \\
	\frac{\partial \delta}{\partial t} &=& v_\gamma.
\end{eqnarray}
Here, $\alpha_g = k_g / (\rho_g C_{p,g})$ is the thermal diffusivity of the vapor, defined by its thermal conductivity $k_g$, density $\rho_g$, and specific heat capacity $C_{p,g}$ at constant pressure. The latent heat of vaporization is represented by $h_{lg}$. The working fluids are water and water vapor, with their exact thermophysical properties summarized in Table~\ref{tab:water}.

\begin{table}[htbp]
	\centering
	\caption{\label{tab:water}Thermophysical properties of water and water vapor at atmospheric pressure.}
	\begin{tabular}{ccc}
		Property & Water (Liquid) & Water (Vapor) \\[0.1cm]
		\hline
		Density $\rho~[\mathrm{kg/m^3}]$ & 958.4 & 0.597\\
		Thermal conductivity $k~[\mathrm{W\cdot m^{-1} K^{-1}}]$ & 0.679 & 0.025\\
		Specific heat capacity $C_{p,g}~[\mathrm{J\cdot kg^{-1} K^{-1}}]$ & 4216 & 2030\\
		Thermal diffusivity $\alpha~[\mathrm{m^2/s}]$ & 1.68e-7 & 2.06e-5\\
		Latent heat of vaporization $h_{lg}~[\mathrm{J/kg}]$ & 2.26e6 &- \\
		Saturation temperature $T_{sat}~[\mathrm{K}]$ & 373.15 &- \\
	\end{tabular}
\end{table}

The exact analytical solution for the interface position and the temperature distribution is given by~\cite{malan2021geometric}:
\begin{eqnarray}
	\delta(t) &=& 2\beta\sqrt{\alpha_g t}, \\
	T(x,t) &=& T_w + \dfrac{T_{sat}-T_w}{\mathrm{erf}(\beta)}\mathrm{erf}\left({\dfrac{x}{2\sqrt{\alpha_gt}}}\right),
\end{eqnarray}
where $\mathrm{erf}$ is the Gaussian error function, and the growth constant $\beta$ is determined by solving the transcendental equation:
\begin{eqnarray*}
	\beta \exp(\beta^2)\mathrm{erf}(\beta)=\dfrac{C_{p,g}(T_w-T_{sat})}{h_{lg}\sqrt{\pi}}.
\end{eqnarray*}

\begin{figure}[htbp]
	\centering
	\includegraphics[width=0.8\textwidth]{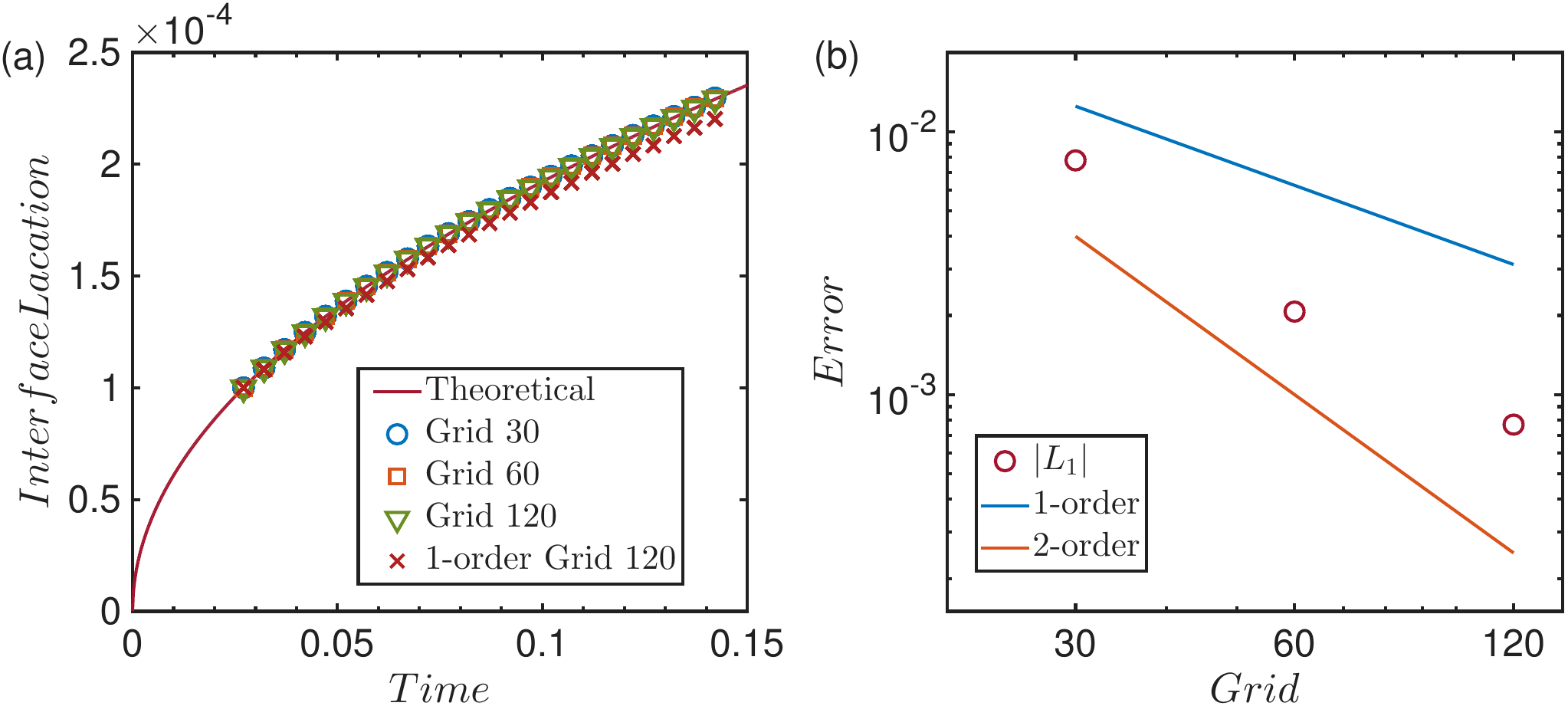}%
	\caption{(a) Interface location versus time for simulations with different grid resolutions, comparing the analytical solution with the proposed method and the baseline method. (b) The spatial order of accuracy, quantified by the $L_1$ error of the temperature distribution inside the vapor layer at the end of the simulations.}
	\label{fig:1dstefan}
\end{figure}

In the present simulations, the wall temperature is set to $T_w = 383.15\,\mathrm{K}$, and the initial vapor layer thickness is $\delta(0) = 0.1\,\mathrm{mm}$ within a total domain length of $1\,\mathrm{mm}$. The temperature profile in the vapor phase is initialized using the analytical solution. Simulations are advanced up to $t = 0.12\,\mathrm{s}$ using three varying grid resolutions: $30$, $60$, and $120$ cells. Figure~\ref{fig:1dstefan}(a) demonstrates good agreement between the numerically predicted interface trajectories and the analytical solution across all grids. Figure~\ref{fig:1dstefan}(b) plots the $L_1$ error norm of the vapor temperature distribution at the final time against the grid spacing, confirming that the proposed method formally achieves second-order spatial accuracy. For comparison, the results obtained using the baseline method (without smooth extension) on the finest grid ($120$ cells) are also included, showing significantly higher numerical errors.

\begin{figure}[htbp]
	\centering
	\includegraphics[width=0.7\textwidth]{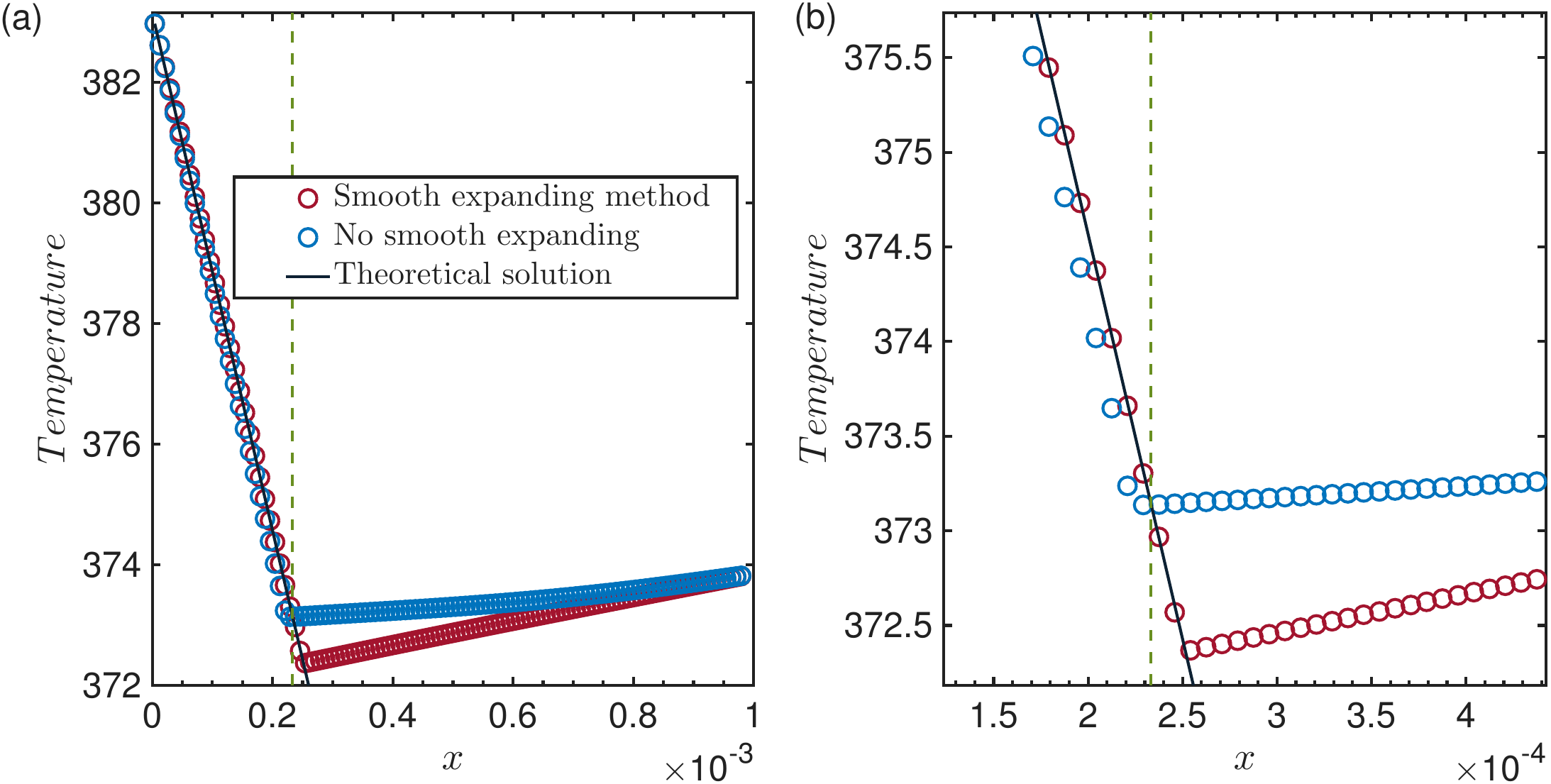}%
	\caption{(a) Comparison of the temperature distributions obtained by the IB methods with and without the smooth extension against the theoretical solution. (b) Magnified view at the interface. The vertical green dashed line marks the theoretical interface location.}
	\label{fig:stefanT}
\end{figure}

To further illustrate the impact of the derivative discontinuity, Figure~\ref{fig:stefanT}(a) compares the temperature profiles obtained using the IB methods with and without the smooth extension. A magnified view near the interface is provided in Fig.~\ref{fig:stefanT}(b). While the baseline method without smooth extension enforces the correct Dirichlet boundary value at the interface, it fails to accurately capture the local temperature gradient. This gradient error propagates into the fluid domain, leading to a noticeable discrepancy between the numerical profile and the analytical solution. By contrast, the proposed smooth extension preserves the gradient continuity at the boundary, resulting in a highly accurate temperature distribution throughout the entire vapor layer.

\subsection{The one-dimensional suction problem}

\begin{figure}[htbp]
	\centering
	\includegraphics[width=0.4\textwidth]{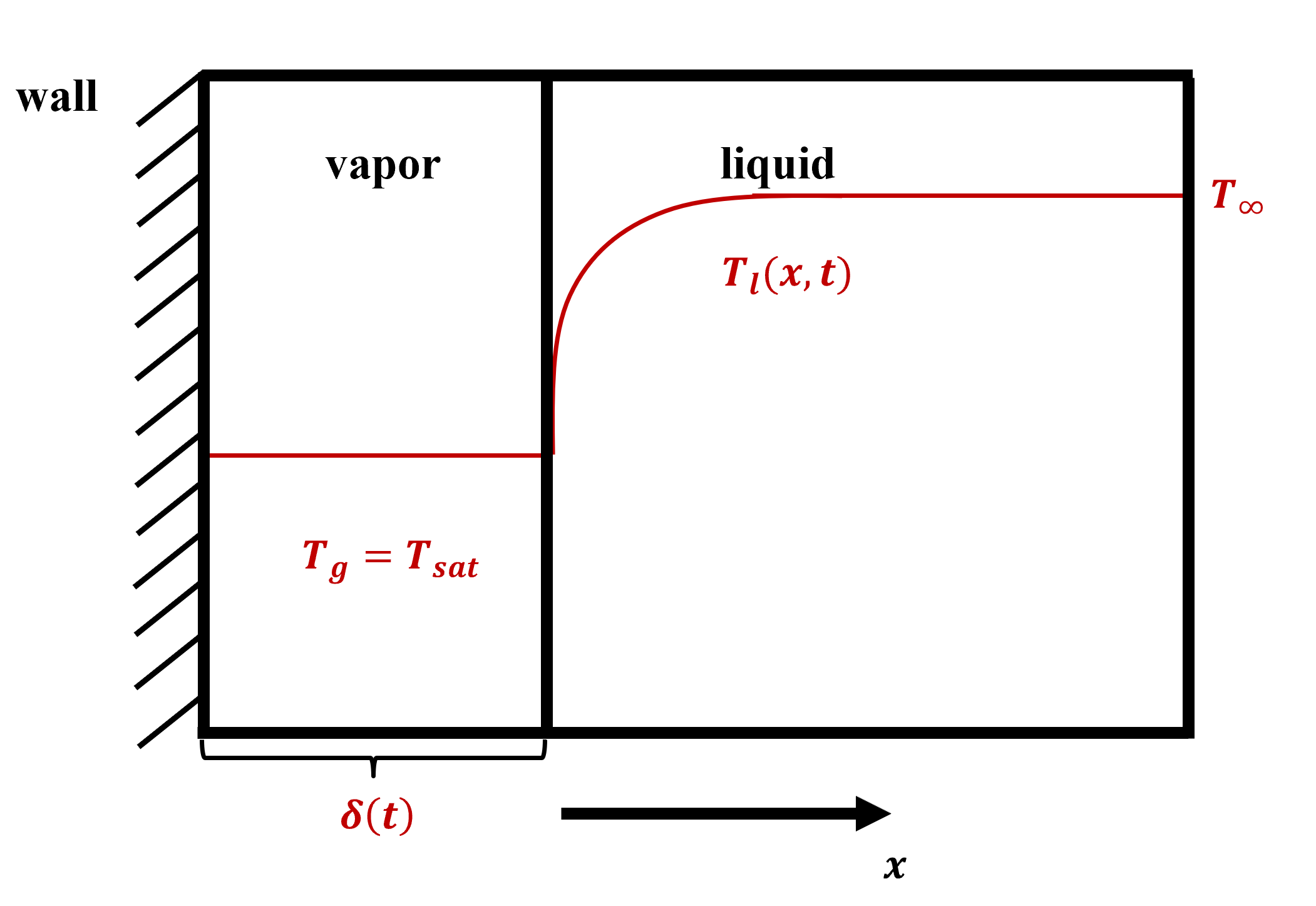}%
	\caption{Configuration of the one-dimensional boiling (suction) problem.}
	\label{fig:sucking}
\end{figure}

The one-dimensional boiling problem, frequently referred to in the literature as the suction problem~\cite{irfan2017front,shao2018computational,welch2000volume}, serves as a more rigorous benchmark than the pure Stefan problem due to the presence of convective heat transfer. As depicted in Fig.~\ref{fig:sucking}, a solid wall is located at $x=0$ and is maintained at the saturation temperature $T_{sat}$. Initially, the liquid phase occupies the semi-infinite space $x > 0$ with a uniform superheated temperature $T_\infty > T_{sat}$. As evaporation proceeds at the wall, a stationary vapor layer of thickness $\delta(t)$ grows, displacing the remaining liquid phase to the right with a uniform velocity $v_l$.

Unlike the Stefan problem, the temperature distribution is solved exclusively within the moving liquid phase ($x > \delta(t)$). The transient convection-diffusion equation is subjected to Dirichlet boundary conditions at both the phase interface and the far-field boundary ($x=L$):
\begin{eqnarray}
	\frac{\partial T}{\partial t} + v_l \frac{\partial T}{\partial x} &=& \alpha_l \frac{\partial^2 T}{\partial x^2}, \qquad \delta(t) < x \le L, \\
	T(\delta(t), t) &=& T_{sat}, \\
	T(L, t) &=& T_\infty.
\end{eqnarray}
The interface dynamics are coupled to the local temperature gradient in the liquid via the Stefan condition:
\begin{eqnarray}
	\rho_l h_{lg} v_\gamma &=& -k_l \left. \frac{\partial T}{\partial x} \right|_{x=\delta(t)^+}, \\
	\frac{d \delta}{d t} &=& v_\gamma.
\end{eqnarray}
By conserving mass across the interface, the uniform liquid advection velocity is given by $v_l = v_\gamma(1-\rho_g/\rho_l)$. Here, the subscript `$l$' denotes the physical quantities of the liquid phase.

The exact analytical solution for this problem is given by~\cite{shao2018computational}:
\begin{eqnarray}
	\delta(t) &=& 2\beta\sqrt{\alpha_g t}, \\
	T(x,t) &=& T_\infty - \left(\frac{T_\infty-T_{sat}}
	{\mathrm{erfc}\left(\beta\frac{\rho_g\sqrt{\alpha_g}}{\rho_l\sqrt{\alpha_l}}\right)}\right)\mathrm{erfc}
	\left(\frac{x}{2\sqrt{\alpha_l t}}+\frac{\beta(\rho_g-\rho_l)}{\rho_g}\sqrt{\frac{\alpha_g}{\alpha_l}}\right),
\end{eqnarray}
where the growth constant $\beta$ is the root of the following transcendental equation:
\begin{equation}
	\exp(\beta^2)\,\mathrm{erf}(\beta)\,
	\left[\beta - \frac{(T_\infty - T_{sat})C_{p,g}k_l\sqrt{\alpha_g}\,
		\exp\left(-\beta^2\dfrac{\rho_g^2\alpha_g}{\rho_l^2\alpha_l}\right)}{h_{lg}k_g\sqrt{\pi\alpha_l}\,
		\mathrm{erfc}\left(\beta\dfrac{\rho_g\sqrt{\alpha_g}}{\rho_l\sqrt{\alpha_l}}\right)}\right]=0.
\end{equation}

The physical properties of the water-vapor system match those listed in Table~\ref{tab:water}. The computational domain length is $L=1\,\mathrm{m}$, with an initial vapor layer thickness of $\delta(0)=0.1\,\mathrm{m}$. The initial liquid temperature profile is prescribed using the analytical solution. The boundary temperatures are fixed at $T_{sat} = T_w = 373.15\,\mathrm{K}$ and $T_\infty = 378.15\,\mathrm{K}$. Simulations are advanced up to $t=3000\,\mathrm{s}$ using four different grid resolutions ranging from $60$ to $180$ cells.

\begin{figure}[htbp]
	\centering
	\includegraphics[width=0.8\textwidth]{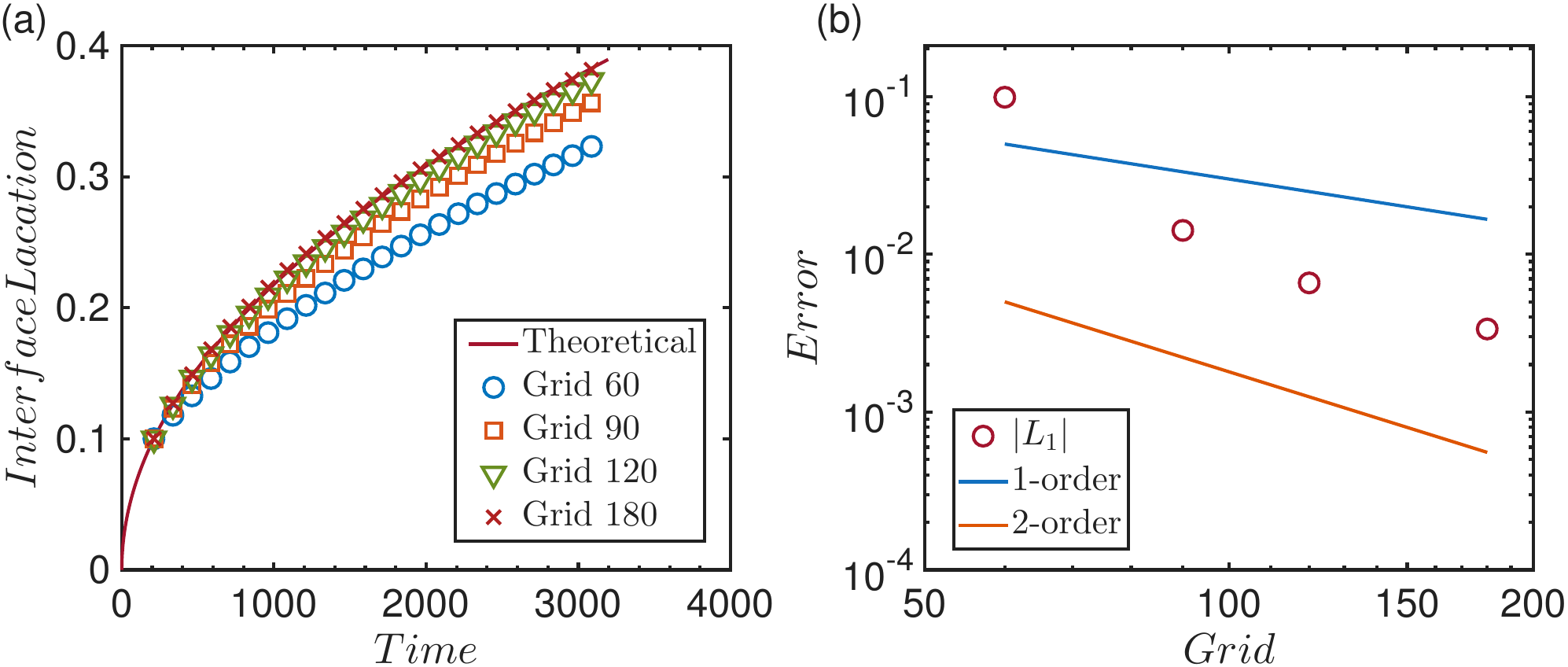}%
	\caption{Numerical results for the one-dimensional boiling problem: (a) temporal evolution of the interface position across different grid resolutions, and (b) spatial order of accuracy quantified by the $L_1$ error of the liquid temperature distribution at $t = 3000\,\mathrm{s}$.}
	\label{fig:suckingaccuracy}
\end{figure}

The tight coupling between the convective term and the moving interface makes this configuration highly sensitive to boundary errors. Figure~\ref{fig:suckingaccuracy}(a) demonstrates the temporal evolution of the interface position. For the refined grids ($120$ and $180$ cells), the numerical predictions closely track the theoretical trajectory. Figure~\ref{fig:suckingaccuracy}(b) illustrates the spatial convergence by plotting the $L_1$ error norm of the final temperature field. Notably, for the coarsest mesh ($60$ cells), initial truncation errors in the interface position rapidly degrade the local temperature gradient. Because the interface velocity is directly driven by this gradient, the errors compound recursively, leading to a substantial deviation from the exact solution. This strong non-linear error amplification explains the apparent super-convergence (exceeding second-order) observed when refining the grid from $60$ to $90$ cells. However, across the adequately resolved grids ($90$, $120$, and $180$ cells), the proposed formulation recovers and maintains the formal second-order spatial accuracy.

\subsection{Three-dimensional bubble growth in superheated liquid}

The growth of a spherical vapor bubble in an infinite pool of superheated liquid~\cite{scriven1959}, serves as a rigorous three-dimensional benchmark for numerical methods handling phase-change interfaces~\cite{kunkelmann2009,sato2013sharp}. Neglecting gravity and assuming the vapor phase remains isothermal at the saturation temperature $T_{sat}$, the bubble expands symmetrically due to evaporation driven by the heat flux from the surrounding superheated liquid ($T_\infty > T_{sat}$).

Similar to the one-dimensional boiling problem, the temperature distribution is solved exclusively within the liquid phase ($r > R(t)$). The three-dimensional transient convection-diffusion equation is subjected to Dirichlet boundary conditions at both the bubble interface and the far-field:
\begin{eqnarray}
	\frac{\partial T_l}{\partial t} + \mathbf{u}_l \cdot \nabla T_l &=& \alpha_l \nabla^2 T_l, \qquad r > R(t), \\
	T_l(R(t), t) &=& T_{sat}, \\
	T_l(r \to \infty, t) &=& T_\infty.
\end{eqnarray}
The radial expansion of the bubble is coupled to the local liquid temperature gradient via the Stefan condition. By evaluating the heat flux at the interface, the growth rate $\dot{R}$ is determined as:
\begin{eqnarray}
	\rho_g h_{lg} \dot{R} &=& k_l \left. \frac{\partial T_l}{\partial r} \right|_{r=R(t)^+}, \\
	\frac{d R}{d t} &=& \dot{R}.
\end{eqnarray}
Due to the density ratio across the phase interface, the expanding vapor drives a radial outward liquid velocity $\mathbf{u}_l$, with its magnitude scaling as $u_r = \dot{R}(1-\rho_g/\rho_l)(R/r)^2$. Here, the subscript `$l$' denotes the physical quantities of the liquid phase.

The exact analytical solution for the bubble radius evolution is given by~\cite{scriven1959}:
\begin{eqnarray}
	R(t) = 2\beta\sqrt{\alpha_l t\,},
\end{eqnarray}
where the dimensionless growth constant $\beta$ is the root of the following transcendental integral equation:
\begin{eqnarray}
	2\beta^2\int_{0}^{1}\exp\left(-\beta^2\left((1-\zeta)^{-2}-2\left(1-\frac{\rho_g}{\rho_l}\right)\zeta-1\right)\right)d\zeta = 
	\frac{\rho_l C_{p,l}(T_{\infty}-T_{sat})}{\rho_g(h_{lg} + (C_{p,l}-C_{p,g})(T_\infty-T_{sat}))}.
\end{eqnarray}
The corresponding analytical radial temperature distribution in the liquid phase reads:
\begin{eqnarray}
	T_l(r,t)&=&T_{\infty} - 2\beta^2\left(\frac{\rho_g(h_{lg}+(C_{p,l}-C_{p,g})(T_\infty-T_{sat}))}{\rho_l C_{p,l}}\right) \nonumber\\
	&& \times\,\int_{1-R(t)/r}^{1} \exp\left(-\beta^2\left((1-\zeta)^{-2}-2\left(1-\frac{\rho_g}{\rho_l}\right)\zeta - 1\right)\right) d\zeta.
\end{eqnarray}

In the present study, the three-dimensional computational domain spans $0.25\,\mathrm{mm} \times 0.25\,\mathrm{mm} \times 0.25\,\mathrm{mm}$. The thermodynamic properties correspond to the water-vapor system evaluated at standard atmospheric pressure. The boundary temperatures are fixed at $T_{sat}=373.15\,\mathrm{K}$ and $T_\infty=378.15\,\mathrm{K}$. The bubble is initially placed at the domain center with a radius of $R(0) = 50\,\mathrm{\mu m}$, and the initial thermal boundary layer thickness is approximately $3.44\,\mathrm{\mu m}$. The initial three-dimensional liquid temperature field is prescribed using the analytical solution. Simulations are advanced up to $t=0.066\,\mathrm{ms}$ using four progressively refined Cartesian grids: $150^3$, $200^3$, $250^3$, and $300^3$ cells. 

\begin{figure}[htbp]
	\centering
	\includegraphics[width=0.7\textwidth]{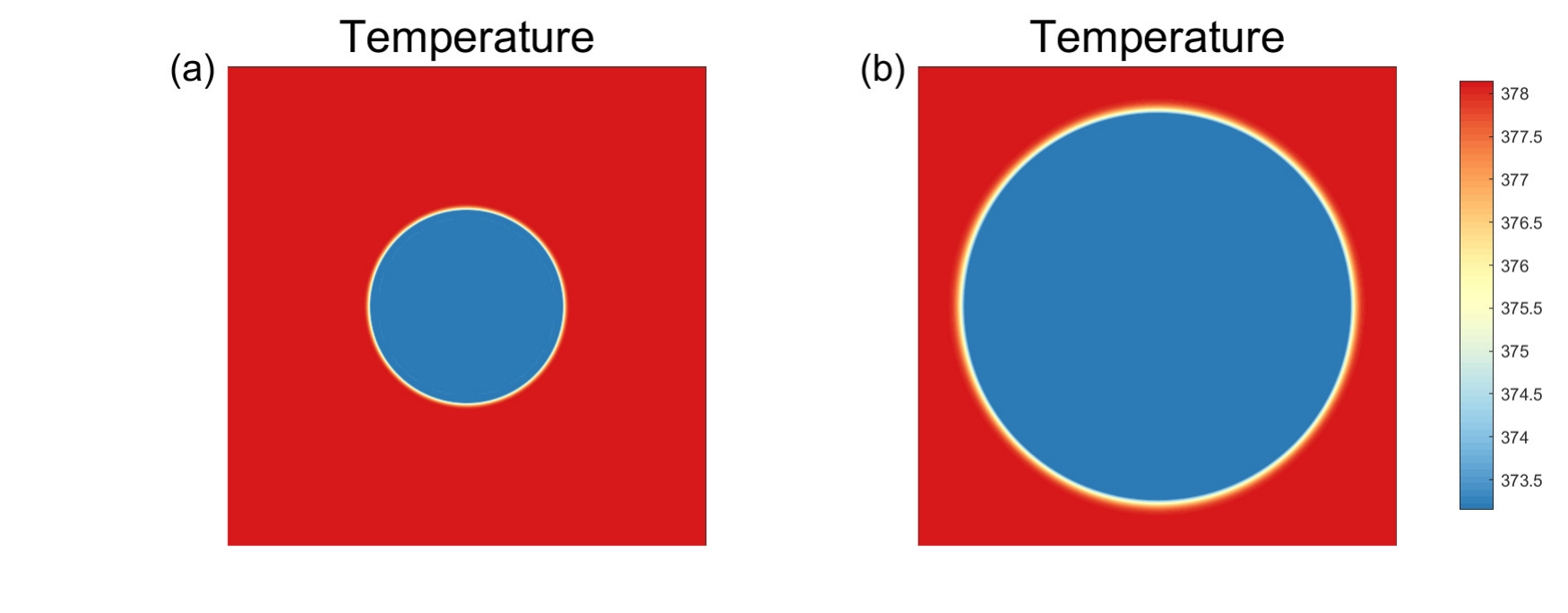}%
	\caption{Mid-plane temperature contours of the three-dimensional bubble growth process at the (a) initial time and (b) final time.}
	\label{fig:3dsukingtime}
\end{figure}

Although the physical problem exhibits spherical symmetry, the simulations are executed on a full unconstrained three-dimensional Cartesian grid. This intentional setup rigorously evaluates the method's capability in resolving severely curved immersed interfaces without invoking symmetry simplifications. Figure~\ref{fig:3dsukingtime} visualizes the mid-plane temperature contours at the initial and final stages of the growth process. 

\begin{figure}[htbp]
	\centering
	\includegraphics[width=0.45\textwidth]{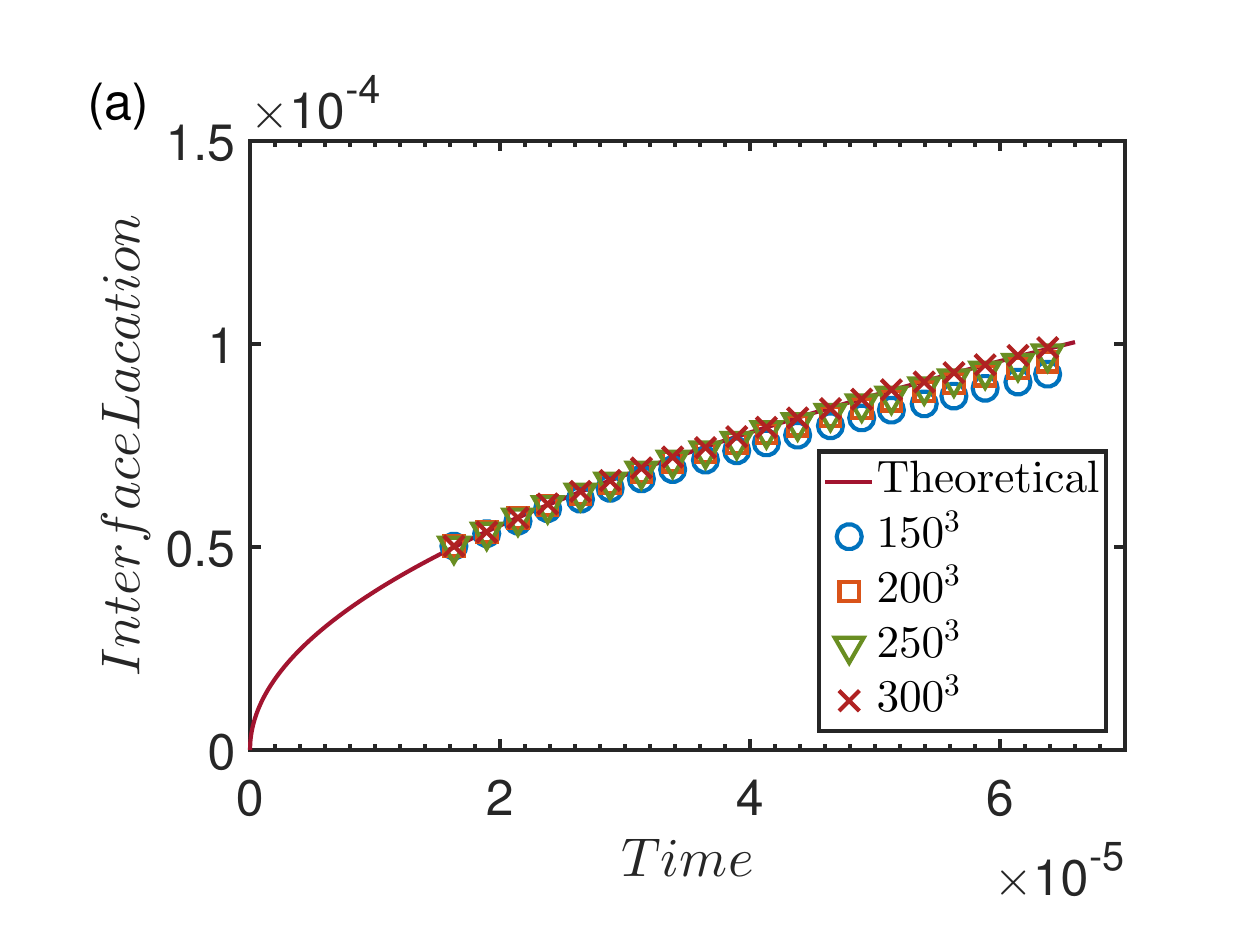}%
	\includegraphics[width=0.45\textwidth]{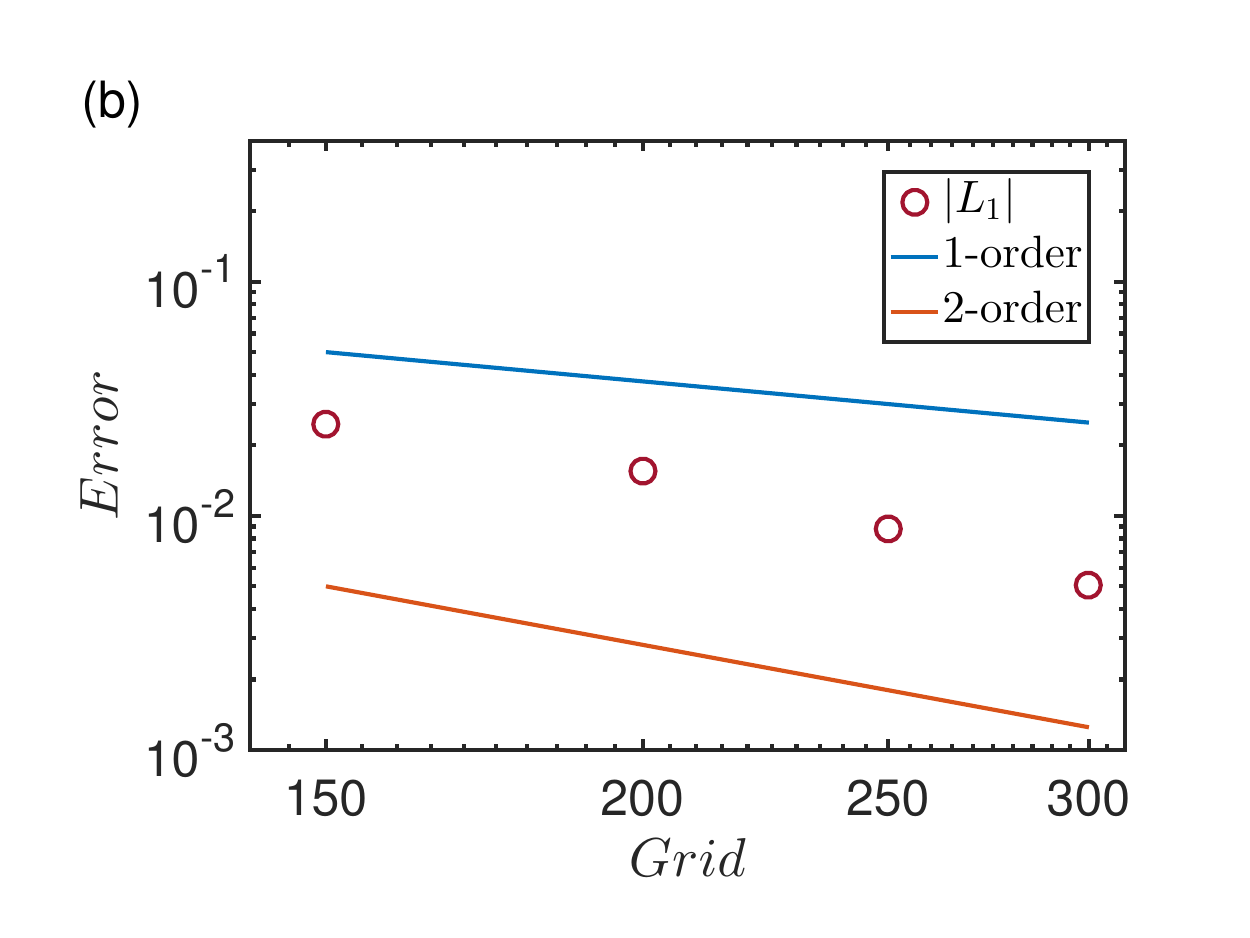}%
	\caption{Numerical results for the three-dimensional bubble growth problem: (a) temporal evolution of the bubble radius across different grid resolutions, and (b) spatial order of accuracy quantified by the $L_1$ error of the liquid temperature distribution at the final time.}
	\label{fig:3dsukingaccuracy}
\end{figure}

As shown in Fig.~\ref{fig:3dsukingaccuracy}(a), the temporal evolution of the numerically predicted bubble radius precisely tracks the analytical solution on the finer grids ($250^3$ and $300^3$). Figure~\ref{fig:3dsukingaccuracy}(b) quantifies the spatial convergence via the $L_1$ error norm of the liquid temperature field at the final time. The error decay trend confirms that the proposed smooth extension strategy successfully maintains formal second-order spatial accuracy, even for highly curved three-dimensional phase interfaces subjected to complex topological coupling.

\subsection{Spontaneous autophoretic motion of an isotropic particle}

The last case considers the spontaneous locomotion of an isotropic phoretic particle~\cite{michelin2013spontaneous,michelin2014phoretic,khair2014expansions,chisholm2016squirmer}. In this configuration, spontaneous symmetry breaking occurs due to the nonlinear coupling between the surface osmotic flow and the advection of the concentration field. The slip velocity at the particle surface is dictated by the local tangential concentration gradient. The governing equations for the fluid and concentration fields are:
\begin{eqnarray}
	&&\nabla \cdot \mathbf{u}=0, \\
	&&\frac{\partial \mathbf{u}}{\partial t} + \mathbf{u}\cdot\nabla\mathbf{u}
	=-\nabla p +\frac{Sc}{Pe}\nabla^2\mathbf{u}, \\
	&&\frac{\partial c}{\partial t}+{\mathbf{u}}\cdot{\nabla c}=\frac{Sc}{Pe}\nabla^2 c,
\end{eqnarray}
with the boundary conditions at the particle surface defined as:
\begin{eqnarray}
	\frac{\partial c}{\partial n}=1,\qquad \mathbf{u}_s=(\mathbf{I}-\mathbf{nn})\cdot\nabla c.
\end{eqnarray}
Here, $Sc=\nu/\kappa$ is the Schmidt number, where $\kappa$ is the mass diffusivity of the solute. The P\'eclet number is defined as $Pe = \frac{AMD}{2\kappa^2}$, with $A$ being the surface emission rate, $M$ the phoretic mobility, and $D$ the particle diameter. Detailed derivations of this physical model can be found in Michelin et al.~\cite{michelin2013spontaneous}. Note that $\mathbf{u}_s$ is the surface slip velocity relative to the particle. 

$\mathbf{U}_c$ and $\mathbf{\Omega}_c$ denote the translational and rotational velocities of the particle, respectively. The actual fluid velocity at the particle surface in the stationary frame of reference is $\mathbf{U}_b=\mathbf{U}_c + \mathbf{\Omega}_c \times (\mathbf{x}_b - \mathbf{X}_c) + \mathbf{u}_s$, where $\mathbf{X}_c$ is the particle centroid and $\mathbf{x}_b$ is the coordinate of the surface point. The full rigid-body dynamics of the particle are governed by Newton-Euler equations~\cite{zhu2024boundary,chen2021instabilities}:
\begin{eqnarray}
	m_p\frac{d\mathbf{U}_c}{dt} &=& \mathbf{F}_h, \\
	\mathbf{I}_p\frac{d\mathbf{\Omega}_c}{dt} &=& \mathbf{T}_h,
\end{eqnarray}
where $m_p$ and $\mathbf{I}_p$ are the mass and the moment of inertia tensor of the particle. $\mathbf{F}_h$ and $\mathbf{T}_h$ are the hydrodynamic force and torque exerted on the particle. 

The hydrodynamic force $\mathbf{F}_h$ and torque $\mathbf{T}_h$ are evaluated by explicitly reconstructing the local hydrodynamic stress tensor, $\boldsymbol{\sigma} = -p\mathbf{I} + \mu(\nabla \mathbf{u} + \nabla \mathbf{u}^T)$, directly at the particle surface. This explicit reconstruction builds upon the original probe method~\cite{vanella2020}. Following the high-order stress recovery principles detailed in Chapter 10 of Verzicco et al.~\cite{verzicco2025introduction}, we extend the baseline approach into a two-point probe scheme. For each surface boundary point $\mathbf{x}_b$, two off-wall probe points are defined along the outward normal $\mathbf{n}$ at distances of $1.5h$ and $2h$ into the fluid domain, denoted as $\mathbf{x}_1 = \mathbf{x}_b + 1.5h\mathbf{n}$ and $\mathbf{x}_2 = \mathbf{x}_b + 2h\mathbf{n}$. The fluid pressure and velocity gradients are first interpolated from the background Eulerian grid to these Lagrangian probe locations. Using the fluid stresses $\boldsymbol{\sigma}_1$ and $\boldsymbol{\sigma}_2$ evaluated at these two probes, the surface stress $\boldsymbol{\sigma}_b$ is determined via linear extrapolation:
\begin{equation}
	\boldsymbol{\sigma}_b = 4\boldsymbol{\sigma}_1 - 3\boldsymbol{\sigma}_2.
\end{equation}
With the local surface stress accurately recovered, the total force and torque exerted on the particle follow from direct surface integration:
\begin{eqnarray}
	\mathbf{F}_h &=& \oint_{S_p} \boldsymbol{\sigma}_b \cdot \mathbf{n} \,dS, \\
	\mathbf{T}_h &=& \oint_{S_p} (\mathbf{x}_b - \mathbf{X}_c) \times (\boldsymbol{\sigma}_b \cdot \mathbf{n}) \,dS.
\end{eqnarray}

The problem is simulated by using the numerical method presented here. A uniform grid with the resolution of $D/\Delta X=24$ is employed for all simulations with different parameters. The lateral section of domain is $20D \times 20D$. The domain length in the swimming direction is $120D$ for $Pe\le7$ and $80D$ for higher $Pe$, respectively. Since at smaller P\'{e}clet numbers, longer distance is needed for the particle to reach the final steady state with velocity $U_\infty$. The numerical results are compared with the analytical solutions in the Stokes regimes~\cite{michelin2013spontaneous}, which requires that the final Reynolds number $Re=U_\infty D / \nu \ll 1$.

It should be noted that when Schmidt number is fixed, the final Reynolds number $Re$ increases with $Pe$. However, in order to keep in the Stokes regime, one must ensure $Re \ll 1$. Therefore, two groups of simulations are conducted in the current validation. For the first group we fix the Schmidt number at $Sc=1$ and gradually increasing $Pe$, which results in an increasing final Reynolds number $Re$. For the second group the P\'{e}clet number is gradually increased from $Pe=2$ to $14$, while $Sc$ is also increases so that the final Reynolds number is fixed at $Re\approx0.1$. 

\begin{figure}
	\centering
	\includegraphics[width=0.8\textwidth]{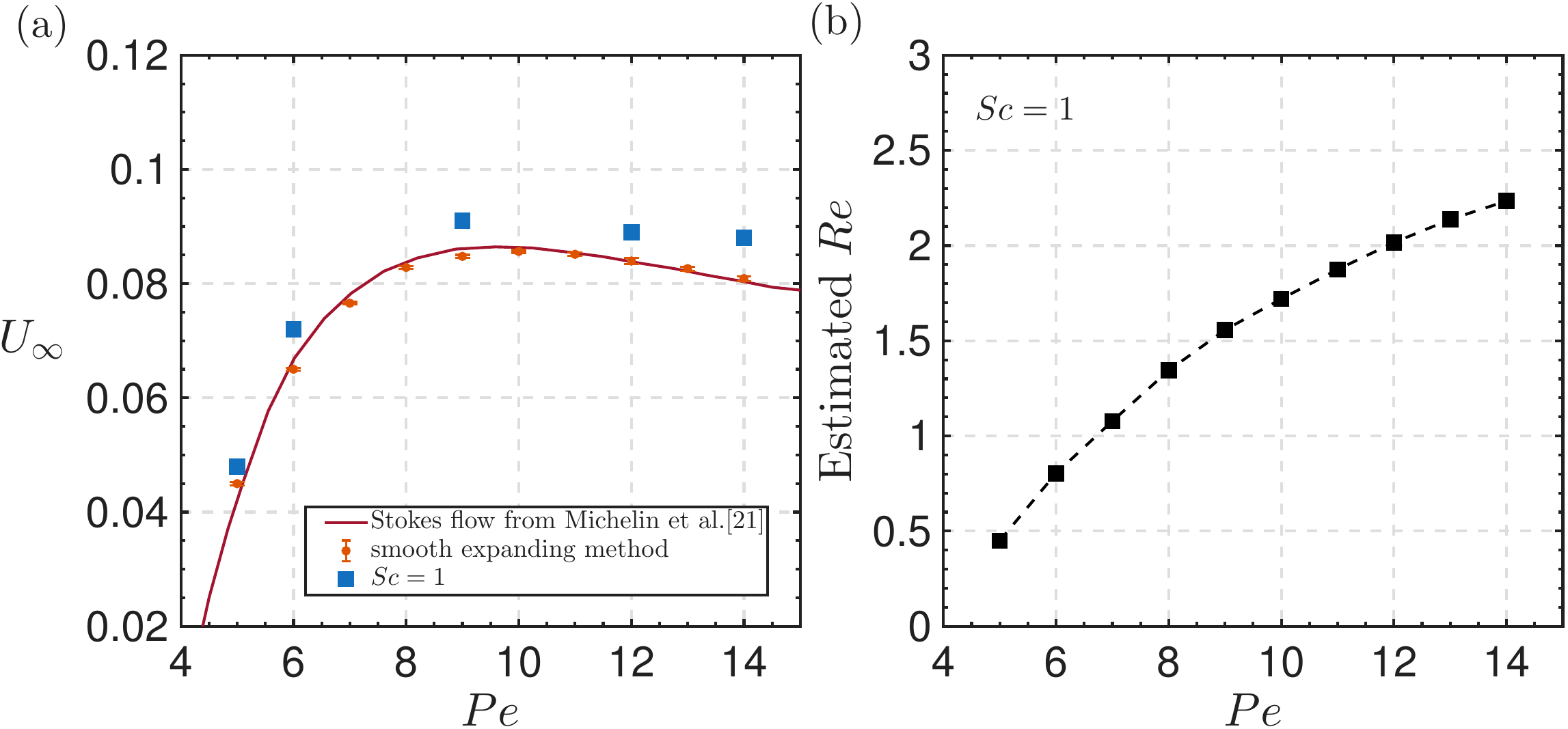}%
	\caption{Steady-state terminal velocity $U_\infty$ as a function of the P\'{e}clet number $Pe$. Our Navier-Stokes simulations (yellow markers with error bars) are compared against the idealized Stokes-flow theoretical baseline~\cite{michelin2013spontaneous} (solid red line) and the fixed $Sc=1$ cases (blue squares). (b) Estimated terminal Reynolds number $Re$ as a function of $Pe$ under the fixed Schmidt number ($Sc=1$) condition.}\label{fig:U-Pe}
\end{figure}

\begin{figure}
	\centering
	\includegraphics[width=0.8\textwidth]{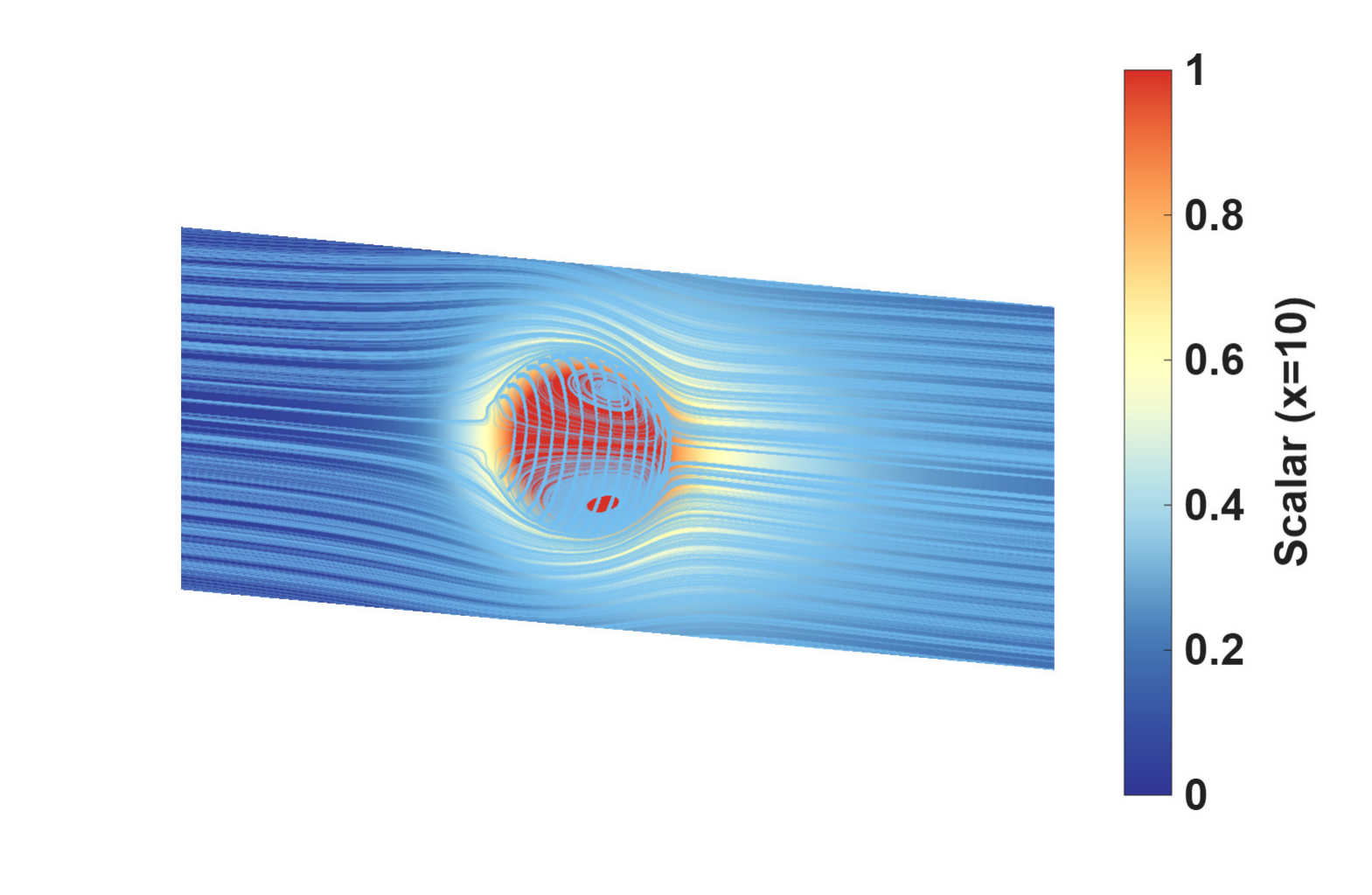}%
	\caption{Flow visualization for $Pe=10$ and $Sc=1$ at $t=400$. The solute concentration field on the plane $x=10$ is shown as a cross-sectional contour map,  overlaid with streamlines in the body-fixed frame computed from the relative velocity $\mathbf{u}-\mathbf{U}_c$ (light-blue curves)}
	\label{fig:Pe10-streamline}
\end{figure}

Figure~\ref{fig:U-Pe}(a) presents the variation of $U_\infty$ versus $Pe$ for the two groups of simulations, and compared with the analytical predictions from~\cite{michelin2013spontaneous}. It can be seen that for the second group with fixed $Re\approx0.1$, the prediction of our numerical method agrees accurately with the theoretical prediction since both numerical and theory are in the Stokes regime. However, the final velocity of the first group with fixed $Sc=1$ is systematically larger than the value of Stokes regime. This is reasonable since the final Reynolds number, as shown in Fig.~\ref{fig:U-Pe}(b), is too high and exceeds the limit for Stokes regime. The flow field for the case with $Pe=10$ and $Sc=1$ is shown in Fig.~\ref{fig:Pe10-streamline}. Figure~\ref{fig:Pe10-streamline} shows the corresponding local streamlines in the boxed region of scalar field, plotted in the body-fixed frame. The solute-induced tangential concentration gradients generate a phoretic slip along the particle surface, leading to a surface-driven shear flow. The shear flow drives a directed motion of the surrounding fluid and sustains the self-propulsion.

\section{Conclusions}

In summary, we developed a smooth extension strategy for the diffuse-interface immersed boundary method within an MLS-IB framework. Traditional diffuse-interface schemes typically suffer from a reduction to first-order accuracy near phase boundaries because the regularized forcing leads to discontinuous first derivatives across the interface, which directly degrades interpolation accuracy. By introducing auxiliary forcing layers on the solid side and reconstructing the scalar values through normal-direction extrapolation, the proposed method enforces continuity of the first-order derivative across the interface and thereby removes the dominant local truncation error source. To avoid introducing artificial mass sources and compromising the projection step in incompressible solvers, this smooth extension is applied only to the scalar field, while the velocity field is treated with the non-iterative and explicit diffuse-interface MLS-IB formulation developed previously in our group.

The numerical framework was systematically validated against classical phase-change benchmarks, including the one-dimensional Stefan evaporation problem and the one-dimensional boiling (suction) problem, as well as three-dimensional spherical bubble growth in a superheated liquid. In these tests, the interface motion is directly coupled to the near-interface scalar gradient via the Stefan condition, making the results highly sensitive to boundary errors. Grid convergence analyses confirmed that the smooth extension obtained nearly second-order spatial accuracy for the scalar field and markedly improves the prediction of interfacial gradients, yielding interface trajectories and temperature fields that closely match the analytical solutions. For the 3D bubble-growth case, the method maintained second-order convergence for a curved moving interface, demonstrating robustness for curved geometries. Furthermore, we simulated the spontaneous locomotion of an isotropic autophoretic particle to demonstrate the method's capability of resolving complex multi-physics couplings. The simulations confirm that, as long as the flow stays in the Stokes regime with sufficiently small Reynolds number, the results of our numerical method agree accurately with the analytical solution for Stokes flows.

It should be pointed out that the current method can only simulate the phase-changing boundary with one side being fluid, since the smooth extension to the extra layer of grids will interfere with the numerical solution of flow field. Also the smooth extension is only applied to the scalar field. Directly modifying the velocity field near the immersed boundary would interfere with the projection step and may violate the divergence-free constraint in incompressible solvers. Developing an incompressibility-preserving extension for the velocity field with higher-order accuracy is therefore highly desired and will be a topic of future works.

\section*{Acknowledgements}
This work is supported by the NSFC Young Scientists Fund (A) under grant No.~12525209 and the NSFC Excellence Research Group Program for `Multiscale Problems in Nonlinear Mechanics' under grant No.~12588201.

\clearpage

\begin{figure}[htbp]
	\centering
	\includegraphics[width=1\textwidth]{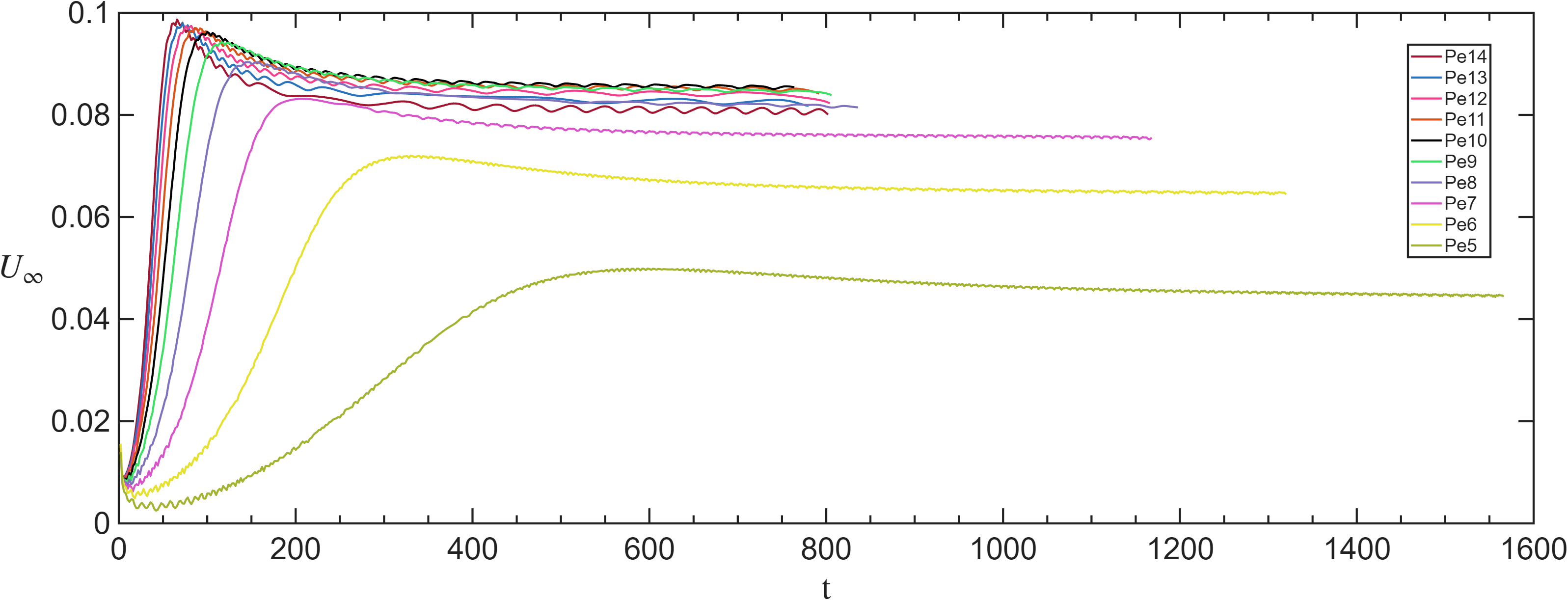}%
	\caption{Transient evolution of the particle swimming velocity over time for P\'eclet numbers ranging from $Pe = 5$ to $14$. The self-propelled motion is initiated by an applied initial perturbation. To systematically isolate the finite-inertia effects, the fluid viscosity in each case is specifically tuned based on the theoretical terminal velocity to maintain a constant target Reynolds number of $Re \approx 0.1$. Following the initial acceleration phase, the swimming velocities for all cases successfully converge to their stable terminal values.}
	\label{fig:U-time}
\end{figure}

\begin{figure}[htbp]
	\centering
	\includegraphics[width=1\textwidth]{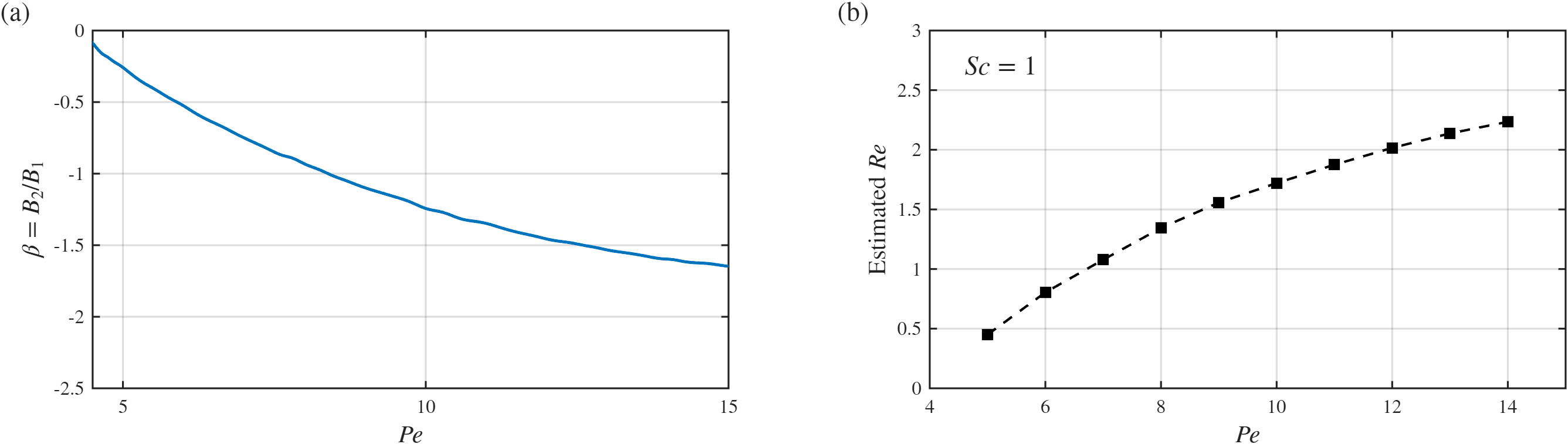}%
	\caption{(a) Steady-state effective squirmer parameter $\beta$ (defined as $B_2/B_1$) as a function of the P\'eclet number $Pe$. The strictly negative values across the entire range indicate a consistent pusher hydrodynamic signature, the magnitude of which increases with $Pe$. (b) Estimated terminal Reynolds number $Re$ as a function of $Pe$ under the fixed Schmidt number ($Sc=1$) condition.}
	\label{fig:beta-Re-Pe}
\end{figure}

\clearpage

\appendix
\section{Derivation of the effective squirmer parameter \texorpdfstring{$\beta$}{beta}}
\label{app:squirmer}

To evaluate the hydrodynamic signature of the autophoretic particle, the effective squirmer parameter $\beta = B_2/B_1$ must be extracted analytically from the dimensionless velocity $U^\infty$ and stresslet $\Sigma^\infty$ defined by Michelin et al.~\cite{michelin2013spontaneous}. This is achieved by mapping the chemically driven slip velocity to the classical Lighthill-Blake squirmer model.

The concentration field on the particle surface is expanded in Legendre polynomials as $c(R, \theta) = \sum_{n=0}^{\infty} c_n P_n(\cos\theta)$, where $c_n$ are the modal amplitudes. The autophoretic slip velocity, driven by the local tangential concentration gradient, is given by $\mathbf{u}_s = M \nabla_s c$. Expressing this in spherical coordinates yields:
\begin{equation}
	\mathbf{u}_s = -M \sum_{n=1}^{\infty} c_n \sin\theta P_n'(\cos\theta) \mathbf{e}_\theta,
\end{equation}
where $P_n'$ denotes the derivative of the Legendre polynomial with respect to its argument.

In the classical squirmer framework~\cite{chisholm2016squirmer}, the tangential surface velocity is prescribed as:
\begin{equation}
	\mathbf{u}_s = \sum_{n=1}^{\infty} \frac{2}{n(n+1)} B_n \sin\theta P_n'(\cos\theta) \mathbf{e}_\theta,
\end{equation}
where $B_n$ are the squirmer modes. Matching the terms of these two expansions directly relates the squirmer modes to the concentration amplitudes:
\begin{equation}
	B_n = -\frac{n(n+1)}{2} M c_n.
\end{equation}

For the first mode ($n=1$), this relation yields $B_1 = -M c_1$. According to Michelin et al.~\cite{michelin2013spontaneous}, the dimensionless swimming speed of an isotropic active particle is derived as $U^\infty = - \frac{2}{3} M c_1$. Substituting $c_1$ into the $B_1$ expression confirms the standard squirmer kinematic relation:
\begin{equation}
	B_1 = \frac{3}{2} U^\infty.
\end{equation}

For the second mode ($n=2$), the general relation gives $B_2 = -3 M c_2$. The dimensionless active stresslet, which governs the leading-order hydrodynamic flow field in the far region, is defined analytically by Michelin et al. as $\Sigma^\infty = -4\pi M c_2$. Substituting the concentration amplitude $c_2$ yields:
\begin{equation}
	B_2 = \frac{3}{4\pi} \Sigma^\infty.
\end{equation}

Taking the ratio of these two modes provides the effective squirmer parameter utilized in the present study to characterize the pusher hydrodynamic signature:
\begin{equation}
	\beta = \frac{B_2}{B_1} = \frac{1}{2\pi} \frac{\Sigma^\infty}{U^\infty}.
\end{equation}

\clearpage

\bibliographystyle{model1-num-names}
\bibliography{ref}

@article{badalassi2003computation,
  title={Computation of multiphase systems with phase field models},
  author={Badalassi, Vittorio E and Ceniceros, Hector D and Banerjee, Sanjoy},
  journal={Journal of computational physics},
  volume={190},
  number={2},
  pages={371--397},
  year={2003},
  publisher={Elsevier}
}

@article{welch2000volume,
  title={A volume of fluid based method for fluid flows with phase change},
  author={Welch, Samuel WJ and Wilson, John},
  journal={Journal of computational physics},
  volume={160},
  number={2},
  pages={662--682},
  year={2000},
  publisher={Elsevier}
}

@article{chisholm2016squirmer,
  title={A squirmer across Reynolds numbers},
  author={Chisholm, Nicholas G and Legendre, Dominique and Lauga, Eric and Khair, Aditya S},
  journal={Journal of Fluid Mechanics},
  volume={796},
  pages={233--256},
  year={2016},
  publisher={Cambridge University Press}
}

@article{chen2021instabilities,
  title={Instabilities driven by diffusiophoretic flow on catalytic surfaces},
  author={Chen, Yibo and Chong, Kai Leong and Liu, Luoqin and Verzicco, Roberto and Lohse, Detlef},
  journal={Journal of Fluid Mechanics},
  volume={919},
  pages={A10},
  year={2021},
  publisher={Cambridge University Press}
}

@book{verzicco2025introduction,
  title={An Introduction to Immersed Boundary Methods},
  author={Verzicco, Roberto and de Tullio, Marco D and Viola, Francesco},
  series={Cambridge Monographs on Applied and Computational Mathematics},
  year={2025},
  publisher={Cambridge University Press},
  address={Cambridge}
}

@article{malan2021geometric,
  title={A geometric VOF method for interface resolved phase change and conservative thermal energy advection},
  author={Malan, LC and Malan, Arnaud G and Zaleski, St{\'e}phane and Rousseau, PG},
  journal={Journal of Computational Physics},
  volume={426},
  pages={109920},
  year={2021},
  publisher={Elsevier}
}

@article{shao2018computational,
  title={A computational framework for interface-resolved DNS of simultaneous atomization, evaporation and combustion},
  author={Shao, Changxiao and Luo, Kun and Chai, Min and Wang, Haiou and Fan, Jianren},
  journal={Journal of Computational Physics},
  volume={371},
  pages={751--778},
  year={2018},
  publisher={Elsevier}
}

@article{sato2013sharp,
  title={A sharp-interface phase change model for a mass-conservative interface tracking method},
  author={Sato, Yohei and Ni{\v{c}}eno, Bojan},
  journal={Journal of Computational Physics},
  volume={249},
  pages={127--161},
  year={2013},
  publisher={Elsevier}
}

@article{peskin1972,
  title={Flow patterns around heart valves: a numerical method},
  author={Peskin, C. S. },
  journal={Journal of Computational Physics},
  year={1972},
  volume={10},
  pages={252--271},
}

@article{ostilla2015multiple,
  title={A multiple-resolution strategy for direct numerical simulation of scalar turbulence},
  author={Ostilla-M{\'o}nico, Rodolfo and Yang, Yantao and Van Der Poel, Erwin P and Lohse, Detlef and Verzicco, Roberto},
  journal={Journal of Computational Physics},
  volume={301},
  pages={308--321},
  year={2015},
  publisher={Elsevier}
}

@article{seo2011sharp,
  title={A sharp-interface immersed boundary method with improved mass conservation and reduced spurious pressure oscillations},
  author={Seo, Jung Hee and Mittal, Rajat},
  journal={Journal of computational physics},
  volume={230},
  number={19},
  pages={7347--7363},
  year={2011},
  publisher={Elsevier}
}

@article{mittal2008versatile,
  title={A versatile sharp interface immersed boundary method for incompressible flows with complex boundaries},
  author={Mittal, Rajat and Dong, Haibo and Bozkurttas, Meliha and Najjar, FM and Vargas, Abel and Von Loebbecke, Alfred},
  journal={Journal of computational physics},
  volume={227},
  number={10},
  pages={4825--4852},
  year={2008},
  publisher={Elsevier}
}

@article{udaykumar2001sharp,
  title={A sharp interface Cartesian grid method for simulating flows with complex moving boundaries},
  author={Udaykumar, HS and Mittal, R and Rampunggoon, P and Khanna, A},
  journal={Journal of computational physics},
  volume={174},
  number={1},
  pages={345--380},
  year={2001},
  publisher={Elsevier}
}

@Inbook{vanella2020,
  author="Vanella, Marcos and Balaras, Elias",
  title="Direct {Lagrangian} Forcing Methods Based on Moving Least Squares",
  bookTitle="Immersed Boundary Method : Development and Applications",
  year="2020",
  publisher="Springer Singapore",
  address="Singapore",
  pages="45--79",
}

@article{khair2014expansions,
  title={Expansions at small Reynolds numbers for the locomotion of a spherical squirmer},
  author={Khair, Aditya S and Chisholm, Nicholas G},
  journal={Physics of Fluids},
  volume={26},
  number={1},
  year={2014},
  publisher={AIP Publishing}
}

@article{kim2001,
  title={An Immersed-Boundary Finite-Volume Method for Simulations of Flow in Complex Geometries},
  author={ Kim, Jungwoo  and  Kim, Dongjoo  and  Choi, Haecheon },
  journal={Journal of Computational Physics},
  volume={171},
  pages={132--150},
  year={2001},
}

@article{michelin2014phoretic,
  title={Phoretic self-propulsion at finite P{\'e}clet numbers},
  author={Michelin, S{\'e}bastien and Lauga, Eric},
  journal={Journal of fluid mechanics},
  volume={747},
  pages={572--604},
  year={2014},
  publisher={Cambridge University Press}
}

@article{mittal2005,
  title={Immersed boundary methods},
  author={Mittal, Rajat and Iaccarino, Gianluca},
  journal={Annu. Rev. Fluid Mech.},
  volume={37},
  pages={239--261},
  year={2005},
}

@article{wang2011,
  title={An immersed boundary method based on discrete stream function formulation for two- and three-dimensional incompressible flows},
  author={ Wang, S.  and  Zhang, X. },
  journal={Journal of Computational Physics},
  volume={230},
  number={9},
  pages={3479--3499},
  year={2011},
}

@article{lai2000,
  title={An Immersed Boundary Method with Formal Second-Order Accuracy and Reduced Numerical Viscosity},
  author={ Lai, M. C.  and  Peskin, C. S. },
  journal={Journal of Computational Physics},
  volume={160},
  number={2},
  pages={705--719},
  year={2000},
}

@article{uhlmann2005,
  title={An immersed boundary method with direct forcing for the simulation of particulate flows},
  author={Uhlmann, Markus},
  journal={Journal of Computational Physics},
  volume={209},
  number={2},
  pages={448--476},
  year={2005},
  publisher={Elsevier}
}

@article{mao2026explicit,
  title={An explicit fully one-sided diffuse-interface immersed boundary method for compressible flows},
  author={Mao, Qian and Cabezas, Jian Eduardo Cardenas and Zhao, Song and Boivin, Pierre and Favier, Julien},
  journal={Journal of Computational Physics},
  pages={114721},
  year={2026},
  publisher={Elsevier}
}

@article{beyer1992analysis,
  title={Analysis of a one-dimensional model for the immersed boundary method},
  author={Beyer, R P and LeVeque, R J},
  journal={SIAM Journal on Numerical Analysis},
  volume={29},
  number={2},
  pages={332--364},
  year={1992},
  publisher={SIAM}
}

@article{stein2016immersed,
  title={Immersed boundary smooth extension: a high-order method for solving PDE on arbitrary smooth domains using Fourier spectral methods},
  author={Stein, D B and Guy, R D and Thomases, B},
  journal={Journal of Computational Physics},
  volume={304},
  pages={252--274},
  year={2016},
  publisher={Elsevier}
}

@article{chen2023,
  title={An explicit and non-iterative moving-least-squares immersed-boundary method with low boundary velocity error},
  author={Chen, W. and Zou, S. and Cai, Q. and Yang, Y.},
  journal={J. Comp. Phys.},
  year={2023},
  volume={474},
  pages={111803},
}

@article{irfan2017front,
  title={A front tracking method for direct numerical simulation of evaporation process in a multiphase system},
  author={Irfan, Muhammad and Muradoglu, Metin},
  journal={Journal of Computational Physics},
  volume={337},
  pages={132--153},
  year={2017},
  publisher={Elsevier}
}

@article{hardt2008,
  title={{Evaporation model for interfacial flows based on a continuum-field representation of the source terms}},
  author={S. Hardt and F. Wondra},
  journal={Journal of Computational Physics},
  volume={227},
  pages={5871--5895},
  year={2008}
}

@article{kunkelmann2009,
  title={{CFD simulation of boiling flows using the volume-of-fluid method within openfoam}},
  author={Christian Kunkelmann and Peter Stephan},
  journal={Numerical Heat Transfer, Part A},
  volume={56},
  pages={631--646},
  year={2009}
}

@article{scriven1959,
  title={{On the dynamics of phase growth}},
  author={Scriven, L.E.},
  journal={Chemical Engineering Science},
  volume={10},
  number={1/2},
  pages={1--18},
  year={1959}
}

@article{zhu2024boundary,
  title={A boundary condition-enhanced direct-forcing immersed boundary method for simulations of three-dimensional phoretic particles in incompressible flows},
  author={Zhu, Xiaojue and Chen, Yibo and Chong, Kai Leong and Lohse, Detlef and Verzicco, Roberto},
  journal={Journal of Computational Physics},
  volume={509},
  pages={113028},
  year={2024},
  publisher={Elsevier}
}

@article{vanella2009,
  title={A moving-least-squares reconstruction for embedded-boundary formulations},
  author={ Vanella, Marcos  and  Balaras, Elias },
  journal={Journal of Computational Physics},
  volume={228},
  number={18},
  pages={6617-6628},
  year={2009},
}

@article{wang2017immersed,
  title={An immersed boundary method for fluid--structure interaction with compressible multiphase flows},
  author={Wang, Li and Currao, Gaetano MD and Han, Feng and Neely, Andrew J and Young, John and Tian, Fang-Bao},
  journal={Journal of Computational Physics},
  volume={346},
  pages={131--151},
  year={2017},
  publisher={Elsevier}
}

@article{souza2022multi,
  title={Multi-phase fluid--structure interaction using adaptive mesh refinement and immersed boundary method},
  author={Souza, Pedro Ricardo C and Neto, H{\'e}lio Ribeiro and Villar, Millena Martins and Vedovotto, Jo{\~a}o Marcelo and Cavalini Jr, Aldemir Ap and Neto, Aristeu Silveira},
  journal={Journal of the Brazilian Society of Mechanical Sciences and Engineering},
  volume={44},
  number={4},
  pages={152},
  year={2022},
  publisher={Springer}
}

@article{yan2024improved,
  title={Improved diffuse interface-immersed boundary method for three-dimensional multiphase fluids--structure interaction with moving contact lines},
  author={Yan, Haoran and Zhang, Guiyong and Wang, Dong and Lu, Yunpeng and Wang, Shuangqiang},
  journal={Applied Ocean Research},
  volume={151},
  pages={104181},
  year={2024},
  publisher={Elsevier}
}

@article{jin2022combined,
  title={A combined volume of fluid and immersed boundary method for modeling of two-phase flows with high density ratio},
  author={Jin, Qiu and Hudson, Dominic and Price, W Geraint},
  journal={Journal of Fluids Engineering},
  volume={144},
  number={3},
  pages={031402},
  year={2022},
  publisher={American Society of Mechanical Engineers}
}

@article{liu2022enthalpy,
  title={Enthalpy-based immersed boundary-lattice Boltzmann model for solid-liquid phase change in porous media under local thermal non-equilibrium condition},
  author={Liu, Xiang and Tong, Zi-Xiang and He, Ya-Ling},
  journal={International Journal of Thermal Sciences},
  volume={182},
  pages={107786},
  year={2022},
  publisher={Elsevier}
}

@article{jost2021direct,
  title={Direct forcing immersed boundary methods: Improvements to the ghost-cell method},
  author={Jost, Antoine Michael Diego and Glockner, St{\'e}phane},
  journal={Journal of Computational Physics},
  volume={438},
  pages={110371},
  year={2021},
  publisher={Elsevier}
}

@article{kempe2012,
  title={An improved immersed boundary method with direct forcing for the simulation of particle laden flows},
  author={Kempe, Tobias and FrHlich, Jochen},
  journal={Journal of Computational Physics},
  volume={231},
  number={9},
  pages={3663--3684},
  year={2012},
}

@article{verzicco1996,
  title={A Finite-Difference Scheme for Three-Dimensional Incompressible Flows in Cylindrical Coordinates},
  author={ Verzicco, R.  and  Orlandi, P. },
  journal={Journal of Computational Physics},
  volume={123},
  number={2},
  pages={402--414},
  year={1996},
}

@article{griffith2005order,
  title={On the order of accuracy of the immersed boundary method: Higher order convergence rates for sufficiently smooth problems},
  author={Griffith, Boyce E and Peskin, Charles S},
  journal={Journal of Computational Physics},
  volume={208},
  number={1},
  pages={75--105},
  year={2005},
  publisher={Elsevier}
}

@article{li1998immersed,
  title={The immersed interface method using a finite element formulation},
  author={Li, Zhilin},
  journal={Applied Numerical Mathematics},
  volume={27},
  number={3},
  pages={253--267},
  year={1998},
  publisher={Elsevier}
}

@article{treeratanaphitak2023diffuse,
  title={Diffuse-interface blended method for imposing physical boundaries in two-fluid flows},
  author={Treeratanaphitak, Tanyakarn and Abukhdeir, Nasser Mohieddin},
  journal={ACS omega},
  volume={8},
  number={17},
  pages={15518--15534},
  year={2023},
  publisher={ACS Publications}
}

@article{michelin2013spontaneous,
  title={Spontaneous autophoretic motion of isotropic particles},
  author={Michelin, S{\'e}bastien and Lauga, Eric and Bartolo, Denis},
  journal={Physics of Fluids},
  volume={25},
  number={6},
  pages={061701},
  year={2013},
  publisher={American Institute of Physics}
}

@article{fadlun2000,
author = {Fadlun, E.A. and Verzicco, R. and Orlandi, P. and Mohd-Yusof, J.},
title = {Combined Immersed-Boundary Finite-Difference Methods for Three-Dimensional Complex Flow Simulations},
year = {2000},
issue_date = {June 10 2000},
publisher = {Academic Press Professional, Inc.},
address = {USA},
volume = {161},
number = {1},
journal = {J. Comput. Phys.},
pages = {35-–60},
}

@article{xiao2022immersed,
  title={Immersed boundary method for multiphase transport phenomena},
  author={Xiao, Wei and Zhang, Hancong and Luo, Kun and Mao, Chaoli and Fan, Jianren},
  journal={Reviews in Chemical Engineering},
  volume={38},
  number={4},
  pages={363--405},
  year={2022},
  publisher={De Gruyter}
}

\end{document}